\documentclass[10pt,a4paper]{article}
\usepackage{jcappub}
\usepackage{pdflscape}
\usepackage{amsmath}
\usepackage{amssymb}
\usepackage{dcolumn}
\usepackage{bm}
\usepackage{color}
\usepackage{epsfig}
\usepackage{amsfonts}
\usepackage{graphicx}
\usepackage{subfigure}
\usepackage{dcolumn}
\usepackage{indentfirst}

\begin{document}

\title{ Black-Bounce in $f(T)$ Gravity}
 
\author[a]{Ednaldo L. B. Junior}
\author[b, c]{Manuel E. Rodrigues} 

\affiliation[a]{Faculdade de Engenharia da Computa\c{c}\~{a}o, 
Universidade Federal do Par\'{a}, Campus Universit\'{a}rio de Tucuru\'{\i}, CEP: 
68464-000, Tucuru\'{\i}, Par\'{a}, Brazil}
\affiliation[b]{Faculdade de Ci\^{e}ncias Exatas e Tecnologia, Universidade Federal do Par\'{a}\\
Campus Universit\'{a}rio de Abaetetuba, CEP 68440-000, Abaetetuba, Par\'{a}, Brazil}
\affiliation[c]{Faculdade de F\'{\i}sica, PPGF, Universidade Federal do 
 Par\'{a}, 66075-110, Bel\'{e}m, Par\'{a}, Brazil}

\emailAdd{ednaldobarrosjr@gmail.com}
\emailAdd{esialg@gmail.com}


\abstract{We study new solutions of black bounce spacetimes formulated in $f(T)$ gravity in four dimensions. First, we present the case of a diagonal tetrad, where a constraint arises in the equations of motion, which is divided into the cases of null torsion, constant torsion, and Teleparallel. The Null Energy Condition (NEC) is still always violated, which implies that the other energy conditions are violated. The solutions are regular in all spacetime and the solution with null torsion exhibits discontinuity between the energy conditions outside and inside the event horizon. Second, we present the case of non-diagonal tetrads. This case is divided into a Simpson-Visser type model and a quadratic model.   The NEC continues to be violated, implying a violation of the other energy conditions. The solutions are regular in all spacetimes. An interesting result is that due to the possibility that the area associated with the metric is different from $4\pi r^2$, the no-go theorem established in the usual $f(T)$ is violated, appearing the new possibility $g_{00}=-g^{11}$, for the components of the metric.} 


\maketitle


\section{Introduction}
\label{sec1}

General Relativity (GR) is the most historically successful theory describing gravitational interaction.  The proof of the bending of starlight passing near the region of the sun, in 1919, showed that RG is very satisfactory in describing local and even cosmological phenomena. However, since 1915, when Karl Schwarzschild obtained the first black hole solution \cite{Sch}, RG has drawn a lot of attention from the scientific community. This type of solution is characterized by a region of space-time where no information beyond a certain limit called the event horizon can return to its point of origin. This region also has a singularity at the origin of the radial coordinate.
\par
After Schwarzschild black hole prediction, which is characterized as a spherically symmetric, static solution in a vacuum, other solutions of fundamental importance to the theory emerged; is the case of the first solution with matter coupling, in this case with Maxwell's electromagnetism, namely the spherically symmetric, static and charged Reissner \cite{Reissner} and Nordstrom \cite{Nordstrom} solution, known as the Reissner-Nordstrom black hole. Another solution, which can be thought of as a generalization of the Schwarzschild solution, is the Kerr (1963) solution \cite{Kerr}, where rotation effects are now accounted for in spherically symmetric space-time. These pioneering works inspired the advancement of black hole research.  The GR has undergone many experimental and observational tests, and with this it is widely accepted by the scientific community.  Recently, the Event Horizon Telescope collaboration, led by John Wardle, presented the first image of a black hole \cite{Horizon}. Research into gravitational waves developed and detected by the LIGO Scientific and Virgo collaboration also has a strong link to this type of structure, although it is not a concrete proof of the existence of these objects due to their similarity to very compact objects, such as neutron stars and white dwarfs.
\par 
However, despite the success, there are some problems that are still open in GR such as the prediction of the space-time singularity, which even though it can be hidden by the event horizon, its presence already represents the violation of physical laws. This singularity is closely linked to the existence of black holes, the origin of the universe, and the quantization of gravitation. The first to present a possible solution to the singularity broblem in the theory was A. Sakharov in a paper on the structure formation of the universe \cite{Sakharov}, later, Bardeen suggests the first spherically symmetric and static solution to the singularity-free GR \cite{bardeen}, but still with the presence of an event horizon in the causal structure. His strategy was to adopt a mass function that depends only on the radial coordinate. This solution became known as a Bardeen regular black hole, thus opening a window for new regular solutions. A numerous class of regular black holes can be found in the literature \cite{39} that in GR are obtained from coupling with Non-Linear Electrodynamics(NED). This type of matter when coupled with gravitation is able to eliminate the space-time singularity via the regular field distribution that covers the central core of the black hole. In this context of non-singular structures {\it Energy Conditions} play an important role in achieving satisfactory results.  First proposed by Hawnking and Ellis to regulate the energy-momentum tensor \cite{HEllis}, the usual energy conditions are: Energy Conditions {\it Null} ($NEC$), {\it Strong} ($SEC$), {\it Dominant} ($DEC$) and {\it Weak} ($WEC$), the violation of each determines the type of fluid of the theory, for example, the Hawking-Penrose \cite{HEllis} singularity theorem requires the $SEC$, and if it is violated, the associated fluid can explain the accelerated expansion of the universe. 
\par 
Another very peculiar type of structure that was extracted from Einstein's field equations are the so-called Wormholes, presented by Michael Morris and Kip Thorne as a transitable structure \cite{Wormholes}, that is, capable of functioning as a tunnel between two distinct and distant regions of the universe, where an individual can freely transit through this tunnel between two points in the universe; more than this, these structures can function as time machines. Later two important works were written on the subject in book form, the first by Visser \cite{Visser} and more recently by, Lobo \cite{LOBO}. It is indispensable that the wormhole solutions present a throat where there is no event horizon; this property imposes a restriction on the type of matter that generates the wormhole, however, for this matter, called exotic matter, the energy-momentum tensor violates the null energy condition of GR.  Wormholes have become even more interesting with the proof of the accelerated expansion of the universe due to the presence of dark energy.  
\par 
The search for singularity-free solutions in GR has led to the recent construction by Simpson and Visser \cite{Visser2}, and later \cite{manuelmarcos}, of a class of regular black holes called the black-bounce, which intersects between the Schwarzschild black hole and the Morris-Thorne wormhole. Between these two solutions, this space-time goes through a black-bounce, an extreme null-bounce, and a traversable wormhole. The regularity of the geometry throughout space is guaranteed by the nonzero bounce parameter, moreover, $r=0$ can be spatial, null, or temporal, characterizing a rather interesting form of a regular black hole, further extending the class of these structures. A non-static evolution version of the spacetime obtained by \cite{Visser2} was investigated in \cite{Visser3} where the metric was rewritten using Eddington-Finklestein coordinates, taking the mass as a function of the null time coordinate, which leads to a structure that interpolates between Vaidya spacetime, a black-bounce and a traversable wormhole. In \cite{Visser4}, a spherically symmetric thin-shell wormhole was considered by combining two black bounce spacetimes via the cut-and-paste method and the analysis of the stability and evolution of this black-bounce wormhole to thin-shell dynamics was performed.
\par 
Although GR is generally accepted, a crucial problem with the theory is that the actual accelerating phase of our universe, shown by the most recent observations of type Ia supernovae \cite{Ia}, can only describe the dynamics of the universe through an exotic fluid, called dark energy, that has negative pressure. As an alternative to this problem one changes the description of the theory through a modification of the Einstein equations. The first attempts to modify GR were in using a Lagrangean density with nonlinear terms in the curvature scalar $R$, known as the Starobinski action \cite{Starobinski}, from which it was generalized to better known and addressed Gravity theory $f(R)$ \cite{fR}, where the action is given in term of an analytic function of the curvature scalar, $S_{f(R)}=\int dx^4\sqrt{-g}[f(R)+2\kappa^2 \mathcal{L}_{Matter}]$. Other modifications are also found, taking the action as a functional of nonlinear terms of a given scalar, the $f(R,\mathcal{T})$ Gravity \cite{fRT} and the $f(\mathcal{G})$ Gravity \cite{fG}.
\par
A large class of regular black hole solutions is found in the literature in the context of modified gravity theories, namely, these solutions for gravity $f(R)$, for example, have been obtained with NED coupling \cite{35marcos} using the metric formalism and characterized by specific parameters that generalize solutions of known regular black holes in GR coupled to NED.  These results were examined for energy conditions and shown that some particular solutions violate only $SEC$, according to \cite{29}. Keeping $f(R)$ and the coupling lagrangian of the NED free, taking only the appropriate choice of the mass function, the analysis of the regularity of the solutions was performed in \cite{47} and demonstrated the existence of two horizons, being an event horizon and another Chauchy horizon and that all the energy conditions are satisfied throughout the space-time, with the exception of $SEC$, which is violated near the Chauchy horizon. On the same path, other regular solutions with NED coupling have now been obtained for a generalized mass function \cite{48}. Other solutions can also be found in \cite{49}.    
\par 
There is also the possibility to describe the gravitational interaction is by considering the inertia of the motion. In this case, we use the torsion of space-time to account for the inertial effects, leaving out the curvature, considering the Reiemann tensor identically zero, thus assuming that the geometry is characterized only by the tetrads as dynamic fields, and the Weitzenbock connection as originating the torsion in space-time, this description of gravitation is able to predict the same global and local phenomena predicted by GR. This theory is known as Teleparallel Theory (TT) \cite{TT} and is dynamically equivalent to GR, where the equations of motion of test particles, are described by a force equation, which can be rewritten as the famous geodesic equation in GR. The scalar constructed in an analogous way to that of curvature is the scalar torsion $T$. Similar to what is done in GR, the direct generalization of this theory is known as $f(T)$ Gravity \cite{fT}, where the action is given in terms of an analytic function of the clarified torsion. Importantly, this generalization arose from a consideration that the action of a theory formulated in this way, could be inspired by NED, such as Born-Infeld (BI). The original work by Ferraro and Fiorini \cite{ferraro1} considers a generalization of TT in which the action is a functional of the clarified torsion with identical dependence to BI, for the scalar $F$.  Solutions of charged black holes were obtained in $f(T)$ Gravity coupled with NED \cite{ednaldo}, which paved the way for obtaining exact solutions, with spherical symmetry, of regular black holes in $f(T)$ theory with the mathematical content of NED, recovering several solutions from GR, for the particular case where $f(T)=T$ \cite{rodrigues4}.  The energy conditions for this theory have been studied in \cite{reboucas} and shown that the inequalities of the energy conditions are different from those of GR, but in the limit $f(T)=T$ are recovered. 
\par
 Then arises the possibility of obtaining black-bounce solutions for $f(T)$ Gravity, where we will initially discuss the solutions of the equations of motion with non-diagonal tetrads and then with the choice of diagonal tetrads. In subsec.\ref{tnula} the case where the scalar torsion is null is analyzed, plotting the behavior of the energy conditions for this solution.  In subsec.\ref{tconstante}, a specific type of black-bounce is presented and the energy conditions of the solution are analyzed. Also at the end of Sec.\ref{secao3} the teleparallel case is briefly explored. For non-diagonal tetrads, in Sec.\ref{secao4}, a constraint is imposed by scalar torsion for regions inside and outside the event horizon. The behavior of the function $f(T)$ are plotted for both cases and a quadratic model of $f(T)$ is explored.

\section{$f(T)$ Gravity}
\label{sec2}
\par
We take a differentiable manifold ``$\mathcal{M}$" where we can define tangent (or co-tangent) space to a point $p$ of that manifold, as the space containing all vectors tangent to it at this point, such that this set of vectors obeys the properties of a vector space. So we call the tangent space $\mathbb{T}_p(\mathcal{M})$. We can generalize this definition to a vector field where the tangent vectors can be defined at any point $p\in \mathcal{M}$, we then represent by $\mathbb{T}(\mathcal{M})$. The space dual to the tangent space is called the co-tangent space $\mathbb{T}^*(\mathcal{M})$ and contains all co-vectors in $\mathcal{M}$. Any element of $\mathbb{T}(\mathcal{M})$ can be written in the form $V=V^{\mu} \partial_{\mu}$ ($\partial_{\mu}=\partial/\partial x^{\mu}$), and every element of $\mathbb{T}^*(\mathcal{M})$ can be written as $\omega=\omega_{\mu} dx^{\mu}$, where $\{\partial_{\mu}\}$ and $\{dx^{\mu}\}$ are the local linearly independent bases of the tangent and co-tangent space respectively.  
\par 
Local bases can relate to general bases by $e_{a}=e^{\;\;\mu}_{a}\partial_{\mu}$ and $e^{a}=e^{a}_{\;\;\mu}dx^{\mu}$, where the matrices satisfy the relations  $e^{a}_{\;\;\mu}e_{a}^{\;\;\nu}=\delta^{\nu}_{\mu}$ and  $e^{a}_{\;\;\mu}e_{b}^{\;\;\mu}=\delta^{a}_{b}$. The general bases of the tangent space satisfy the following commutation relation
\begin{eqnarray}
[e_{a},e_{b}]=f^{c}_{\;\;bc}e_{c}\,,\label{cr1}
\end{eqnarray}
where $f^{c}_{\;\;bc}e_{c}$ are the coefficients of structure or anholonomy.  Using the first equation of Cartan structure  ($de^{c}=-(1/2)f^{c}_{\;\;ab}e^{a}\wedge e^{b}$) and the relationship between the local base, we have $f^{c}_{\;\;ab}=e_{a}^{\;\;\mu}e_{b}^{\nu}(\partial_{\nu}e^{c}_{\;\;\mu}-\partial_{\mu}e^{c}_{\;\;\nu})$. With this, we define an inertial frame when the structure coefficients are all identically zero, i.e. $f^{c}_{\;\;ab}\equiv 0$. In general, for any given frame, we do not necessarily have this condition.  
\par 
With this structure, the row element in $\mathcal{M}$, can be represented in local or general bases as
\begin{eqnarray}
dS^2=g_{\mu\nu}dx^{\mu}dx^{\nu}=\eta_{ab} e^{a}e^{b}\label{ele}\;,\\
e^{a}=e^{a}_{\;\;\mu}dx^{\mu}\;,\;dx^{\mu}=e_{a}^{\;\;\mu} e^{a}\label{the}\;,
\end{eqnarray}
where $g_{\mu\nu}$ is the metric of the space-time, $\eta_{ab}$ the Minkowski metric. The root of the determinant of the metric is given by $\sqrt{-g}=det[e^{a}_{\;\;\mu}]=e$. For a special case of this structure, we can require that the Riemann tensor be identically zero, which is obtained with the Weitzenbok connection  
\begin{eqnarray}
\Gamma^{\sigma}_{\;\;\mu\nu}=e_{a}^{\;\;\sigma}\partial_{\nu}e^{a}_{\;\;\mu}=-e^{a}_{\;\;\mu}\partial_{\nu}e_{a}^{\;\;\sigma}\label{co}\; .
\end{eqnarray}
\par
Using \eqref{co} we can define the components of the Torsion and Contortion tensors 
\begin{eqnarray}
T^{\sigma}_{\;\;\mu\nu}&=&\Gamma^{\sigma}_{\;\;\nu\mu}-\Gamma^{\sigma}_{\;\;\mu\nu}=e_{a}^{\;\;\sigma}\left(\partial_{\mu} e^{a}_{\;\;\nu}-\partial_{\nu} e^{a}_{\;\;\mu}\right)\label{tor}\;,\\
K^{\mu\nu}_{\;\;\;\;\alpha}&=&-\frac{1}{2}\left(T^{\mu\nu}_{\;\;\;\;\alpha}-T^{\nu\mu}_{\;\;\;\;\alpha}-T_{\alpha}^{\;\;\mu\nu}\right)\label{cont}\; .
\end{eqnarray}
\par
Now, using \eqref{tor} and \eqref{cont} we define the tensor $S_{\alpha}^{\;\;\mu\nu}$ in order to simplify the equations of motion,
\begin{eqnarray}
S_{\alpha}^{\;\;\mu\nu}=\frac{1}{2}\left( K_{\;\;\;\;\alpha}^{\mu\nu}+\delta^{\mu}_{\alpha}T^{\beta\nu}_{\;\;\;\;\beta}-\delta^{\nu}_{\alpha}T^{\beta\mu}_{\;\;\;\;\beta}\right)\label{s}\;.
\end{eqnarray}
We can define the scalar of the theory, the scalar torsion:
\begin{equation}
T=T^{\alpha}_{\;\;\mu\nu}S_{\alpha}^{\;\;\mu\nu}\label{t1}\,.
\end{equation}
\par 
We will now establish the equations of motion for the case of a generalized matter source, which we will specify in appropriate time. Using the following Lagrangean density
\begin{eqnarray}
\mathcal{L}=e\left[f(T)+2\kappa^2\mathcal{L}_{m}\right],\label{lagrangean}
\end{eqnarray} 
where $\mathcal{L}_{m}$ is the source of matter and $\kappa^2=8\pi G/c^4$, where $G$ is the Newtonian constant and ``$c$'' the speed of light. We establish the equations of motion using the Euler-Lagrange equations. Taking the derivations with respect to the tetrads
\begin{eqnarray}
&&\frac{\partial\mathcal{L}}{\partial e^{a}_{\;\;\mu}}=f\,ee_{a}^{\;\;\mu}+ef_{T}4e^{a}_{\;\;\alpha}T^{\sigma}_{\;\;\nu\alpha}S_{\sigma}^{\;\;\mu\nu}+2\kappa^2\frac{\partial\mathcal{L}_{m}}{\partial e^{a}_{\;\;\mu}}\,\\
&&\partial_{\alpha}\left[\frac{\partial\mathcal{L}}{\partial (\partial_{\alpha}e^{a}_{\;\;\mu})}\right]=-4f_{T}\partial_{\alpha}\left(ee_{a}^{\;\;\sigma}S_{\sigma}^{\;\;\mu\nu}\right)-4ee_{a}^{\;\;\sigma}S_{\sigma}^{\;\;\mu\gamma}\partial_{\gamma}T f_{TT}+2\kappa^2\partial_{\alpha}\left[\frac{\partial\mathcal{L}_{m}}{\partial (\partial_{\alpha}e^{a}_{\;\;\mu})}\right]\,,
\end{eqnarray}
with $f=f(T)$, $f_{T}=df(T)/dT$ e $f_{TT}=d^2f(T)/dT^2$. With the above expressions and the Euler-Lagrange equations
\begin{eqnarray}
\frac{\partial\mathcal{L}}{\partial e^{a}_{\;\;\mu}}-\partial_{\alpha}\left[\frac{\partial\mathcal{L}}{\partial (\partial_{\alpha}e^{a}_{\;\;\mu})}\right]=0\label{ELeq}\,,
\end{eqnarray} 
and multiplying everything by $e^{-1}e^{a}_{\;\;\beta}/4$, we have the equations of motion for $f(T)$ gravity written as
\begin{eqnarray}
S_{\beta}^{\;\;\mu\alpha}\partial_{\alpha}T f_{TT}+\left[e^{-1}e^{a}_{\;\;\beta}\partial_{\alpha}\left(ee_{a}^{\;\;\sigma}S_{\sigma}^{\;\;\mu\alpha}\right)+T^{\sigma}_{\;\;\nu\beta}S_{\sigma}^{\;\;\mu\nu}\right]f_{T}+\frac{1}{4}\delta^{\mu}_{\beta}f=\frac{\kappa^2}{2}\Theta^{\;\;\mu}_{\beta}\,,\label{meq}
\end{eqnarray} 
where $\Theta_{\beta}^{\;\;\mu}$ is the energy-momentum tensor of the matter source
\begin{eqnarray} 
\Theta_{\beta}^{\;\;\mu}=e^{-1}e^{a}_{\;\;\beta}\left(\frac{\partial\mathcal{L}_{m}}{\partial e^{a}_{\;\;\mu}}-\partial_{\alpha}\frac{\partial\mathcal{L}_{m}}{\partial(\partial_{\alpha}e^{a}_{\;\;\mu})}\right) \label{emtensor}\,.
\end{eqnarray} 
\par
Using the identity
\begin{eqnarray}
e^{-1}e^{a}_{\;\;\nu}\partial_{\alpha}\left(e e_{a}^{\;\;\beta}S_{\beta}^{\;\;\alpha\sigma}\right)+T^{\alpha}_{\;\;\beta\nu}S_{\alpha}^{\;\;\beta\sigma}=\frac{1}{2}\left(G_{\nu}^{\;\;\sigma}-\frac{1}{2}\delta^{\sigma}_{\nu}T\right)\,,
\label{id1}
\end{eqnarray}
where $G^{\;\;\nu}_{\sigma}$ is the Einstein tensor. We can rewrite the equations \eqref{meq} in terms of the effective energy-momentum tensor $\Theta^{(eff)\;\mu}_{\;\;\;\;\;\;\;\nu}$ 
\begin{eqnarray}   
G^{\;\;\mu}_{\nu}=\kappa^2\Theta^{(eff)\;\mu}_{\;\;\;\;\;\;\;\nu}=\frac{2}{f_T}\left[\frac{1}{2}\kappa^2\Theta_{\nu}^{\;\;\mu}-\frac{1}{4}\delta^{\mu}_{\nu}(f-f_T T)-f_{TT}S_{\nu}^{\;\;\mu\alpha}\partial_{\alpha}T\right]\,.\label{Thetaeff}
\end{eqnarray}
\par
As defined, the quantity $\Theta^{(eff)\;\mu}_{\;\;\;\;\;\;\;\nu}$ represents the effective energy-momentum tensor arising from $f(T)$ gravity, which acts as the effective source in the Eeinstein equations. It contains the canonical energy-momentum tensor $\Theta_{\nu}^{\;\;\mu}$ of matter fields, balanced by $f_T^{-1}$ and the additional contributions due to the presence of the $f(T)$ function in the Lagrangean density. 
\par
Based on these changes to the source terms of the Einstein equations, the concept of energy conditions can be extended to modified gravity $f(T)$ when we assume that the matter content of the universe behaves like an anisotropic fluid. With that, we define the effective density and pressures respectively as $\rho^{(eff)}=\Theta^{(eff)\;0}_{\;\;\;\;\;\;\;0}$ e $p_{r}^{(eff)}=-\Theta^{(eff)\;1}_{\;\;\;\;\;\;\;1}$ e $p_t^{(eff)}=-\Theta^{(eff)\;2}_{\;\;\;\;\;\;\;2}$ and therefore the $NEC$, $WEC$, $SEC$ and $DEC$ are,
\begin{eqnarray}
NEC_{1,2}=SEC_{1,2}=WEC_{1,2}:&&\;\;\;\rho^{(eff)}+p_{r,t}^{(eff)}\geq 0\,,\label{NEC}\\
SEC_3:&&\;\;\;\rho^{(eff)}+p_{r}^{(eff)}+2p_t^{(eff)}\geq 0\,,\label{SEC}\\
DEC_{1,2}:&&\;\;\;\rho^{(eff)}-\mid p_{r,t}^{(eff)}\mid \geq 0\,\Longleftrightarrow \rho^{(eff)}\pm p_{r,t}^{(eff)}\geq 0\,,\label{DEC}\\
DEC_{3}=WEC_3:&&\;\;\;\rho^{(eff)}\geq 0\,.\label{WEC}
\end{eqnarray}
Here the indices $\lbrace 1, 2\rbrace$ are associated with $\lbrace$radial, tangential$\rbrace$.

\section{Black-Bounce solution with diagonal tetrada}\label{secao3}

\par

Because of the relation in (\ref{ele}), we can choose several frames that fall into a spherically symmetric and static metric in spherical coordinates. Also the several tetrads matrices will be connected by means of a Lorentz transformation. Taking a diagonal matrix $[\bar{e}^{a}_{\;\;\mu}]=diag[e^{a(r)},e^{b(r)},\Sigma(r),\Sigma(r)\sin\theta]$, to be the dynamic field of the theory, where $a(r), b(r)$ and $\Sigma(r)$ are functions of radial coordinate and are not time-dependent.  Using \eqref{ele} we can reconstruct the metric as being, 
\begin{equation}
dS^2=e^{a(r)}dt^2-e^{b(r)}dr^2-\Sigma(r)^2\left[d\theta^{2}+\sin^{2}\left(\theta\right)d\phi^{2}\right]\label{ltb}\,,
\end{equation}
the determinant of the metric is given by $g=-e^{a(r)+b(r)}\Sigma(r)^4\sin^2\theta$.  
\par
With this, we can now calculate all the geometric objects established in the theory. The non-zero components of the torsion tensor (\ref{tor}) are
\begin{eqnarray}
T^{0}_{\;\;10}=\frac{a'}{2}\,,\;\;\;T^{2}_{\;\;21}=T^{3}_{\;\;31}=-\frac{\Sigma'}{\Sigma}\,\,,\nonumber\\
T^{0}_{\;\;01}=-\frac{a'}{2}\,,\;\;\; T^{2}_{\;\;12}=T^{3}_{\;\;13}=\frac{\Sigma'}{\Sigma}\,,\nonumber\\
T^{3}_{\;\;23}=\cot\theta\,,\;\;\; T^{3}_{\;\;32}=-\cot\theta\,, \label{tt}
\end{eqnarray}
where $'$ represents the derivative with respect to the coordinate $r$, while the non-null components of the contorsion tensor read
\begin{eqnarray}
K_{\;\;\;\;0}^{10}=\frac{a'e^{-b}}{2}\,,\;\;\;K_{\;\;\;\;1}^{22}=K_{\;\;\;\;1}^{33}=-\frac{e^{-b}\Sigma'}{\Sigma}\,\,,\nonumber\\
K_{\;\;\;\;1}^{00}=-\frac{a'e^{-b}}{2}\,,\;\;\;K_{\;\;\;\;2}^{12}=K_{\;\;\;\;3}^{13}=\frac{\Sigma'e^{b}}{\Sigma}\,,\nonumber\\
K_{\;\;\;\;2}^{33}=-\frac{\cot\theta}{\Sigma^2}\,,\;\;K_{\;\;\;\;3}^{23}=\frac{\cot\theta}{\Sigma^2}\,.\label{ctt}
\end{eqnarray}
We also calculate the non-zero components of the tensor $S_{\alpha}^{\;\;\mu\nu}$, giving
\begin{eqnarray}
S_{0}^{\;\;01}=-\frac{e^{-b}\Sigma'}{\Sigma}\,,\;\;S_{0}^{\;\;02}=S_{1}^{\;\;12}=-\frac{\cot\theta}{2\Sigma^2}\,,\;\;S_{0}^{\;\;10}=\frac{e^{-b}\Sigma'}{\Sigma}\,,\;\;
S_{0}^{\;\;20}=S_{1}^{\;\;21}=\frac{\cot\theta}{2\Sigma^2}\,,\nonumber\\
S_{2}^{\;\;12}=S_{3}^{\;\;13}=\frac{e^{-b}(\Sigma a'+2\Sigma')}{4\Sigma}\,,\;\; S_{2}^{\;\;21}=S_{3}^{\;\;31}=-\frac{e^{-b}(\Sigma a'+2\Sigma')}{4\Sigma}.\label{st}
\end{eqnarray}
\par
From the definition of the torsion scalar (\ref{t1}), one gets
\begin{equation}
T=\frac{2e^{-b}\Sigma'\left(\Sigma a'+\Sigma'\right)}{\Sigma^2}  \label{te1}\,.
\end{equation}
We note here that in general, the scalar torsion is an arbitrary function of the radial coordinate $r$.
\par
In order to get a consistent solution, we take the energy-momentum tensor to be, $\Theta_{\nu}^{\;\;\mu}=diag\left[\rho(r), -p_r(r), -p_t(r), -p_t(r)\right]$, where $\rho(r)$, $p_r(r)$ and $p_t(r) $ are the density, radial pressure and tangential pressure, respectively. With that, taking the metric \eqref{ltb} and using \eqref{tt}, \eqref{ctt}, \eqref{st} and \eqref{te1} we can rewrite the equations of motion as,
\begin{eqnarray}
&&-\frac{e^{-b}f_T}{2\Sigma^2}\left(e^{b}-2(\Sigma')^2+\Sigma(b'\Sigma'-a'\Sigma'-2\Sigma'')\right)-\frac{e^{-b}f_{TT} T'\Sigma'}{\Sigma}+\frac{f}{4}=\frac{1}{2}\kappa^2\rho\,,\label{eq1}\\
&&\frac{e^{-b}f_T}{2\Sigma^2}\left(e^{b}-2\Sigma a'\Sigma'-2(\Sigma')^2\right)+\frac{f}{4}=-\frac{1}{2}\kappa^2p_r\,,\label{eq2}\\
&&\frac{\cot\theta f_{TT}T'}{2\Sigma^2}=0\,,\label{eq3}\\
&&-\frac{e^{-b}f_T}{8\Sigma^2}\left[4(\Sigma')^2+\Sigma^2\left((a')^2-a'b'+2a''\right)-\Sigma\left(2b'\Sigma'-6 a'\Sigma'-4\Sigma''\right)\right]\nonumber\\
&&-\frac{e^{-b}f_{TT}T'}{4\Sigma}\left(\Sigma a'+2\Sigma'\right)+\frac{f}{4}=-\frac{1}{2}\kappa^2p_t\,. \label{eq4}
\end{eqnarray}
\par 
Using \eqref{Thetaeff} we can calculate the effective density and radial and tangential effective pressures for regions outside the event horizon, this means $e^{-b}>0$ with the $t$ coordinate is timelike, therefore, 
\begin{eqnarray}
&&\rho^{(eff)}=\frac{1}{2f_T\kappa^2}\left[2\kappa^2\rho+f_T T-f+\frac{4e^{-b}f_{TT}T'\Sigma'}{\Sigma}\right]\,,\label{roeff}\\
&&p_r^{(eff)}=\frac{1}{2\kappa^2f_T}\Big[2\kappa^2 p_r+f-f_T T\Big]\,,\label{preff}\\
&&p_t^{(eff)}=\frac{1}{2\kappa^2f_T}\Bigg[2\kappa^2 p_t-f_T T+f-\frac{e^{-b}f_{TT}T'\left(\Sigma a'+2\Sigma' \right)}{\Sigma}\Bigg]\,.\label{pteff}
\end{eqnarray} 
With this, the energy conditions \eqref{NEC}-\eqref{WEC}, outside the event horizon, are now written as,
\begin{eqnarray}
&&NEC_1=SEC_1=WEC_1=\frac{1}{f_T}\left[\rho+p_r+\frac{2e^{-b}f_{TT}T'\Sigma'}{\kappa^2\Sigma}\right]\geqslant 0\,,\label{NEC1}\\
&&NEC_2=SEC_2=WEC_2=\rho+p_t-\frac{e^{-b}f_{TT}T'\left(\Sigma a'-2\Sigma'\right)}{2\kappa^2\Sigma}\geqslant 0\,,\label{NEC2}\\
&&SEC_3=\rho+p_r+2p_t+\frac{e^{-b}f_{TT}a'T'+f-T}{\kappa^2}\geqslant 0\,,\label{SEC3}\\
&&DEC_1=\rho-p_r+\frac{2e^{-b}f_{TT}T'\Sigma'T-f}{\kappa^2}\geqslant 0\,,\label{DEC1}\\
&&DEC_2=\frac{1}{\kappa^2}\left[2\left(T+\kappa^2(\rho-p_t)\right)+\frac{e^{-b}f_{TT}T'\left(\Sigma a'+6\Sigma'\right)}{\Sigma}-2f\right]\geqslant0\,,\label{DEC2}\\
&&DEC_3=WEC_3=\frac{1}{2\kappa^2f_T}\left[2\kappa^2\rho+\frac{4e^{-b}f_{TT}T'\Sigma''}{\Sigma}+f_TT-f\right]\geqslant0\,.\label{DEC3}
\end{eqnarray}
Note that if $f=T$, for any $\lbrace a(r), b(r)\rbrace$, we recover the energy conditions from General Relativity. This case will be explored further below. For regions inside the event horizon, that is, when the time coordinate $t$ is of a spacelike, with $e^{-b}<0$, we have for effectivy density and pressures,  $\rho^{(eff)}=\Theta^{(eff)\;1}_{\;\;\;\;\;\;\;1}$, $p_{r}^{(eff)}=-\Theta^{(eff)\;0}_{\;\;\;\;\;\;\;0}$ and $p_t^{(eff)}=-\Theta^{(eff)\;2}_{\;\;\;\;\;\;\;2}$, therefore, the energy conditions for such a region are, 
\begin{eqnarray}
&&NEC_1=SEC_1=WEC_1=\frac{1}{f_T}\left[p_r+\rho-\frac{2e^{-b}f_{TT}T'\Sigma'}{\kappa^2\Sigma}\right]\geqslant 0\,,\label{NEC1d}\\
&&NEC_2=SEC_2=WEC_2=\frac{1}{2f_T}\left[2\left(p_t+\rho\right)-\frac{e^{-b}f_{TT}T'(\Sigma a'+2\Sigma')}{\kappa^2\Sigma}\right]\geqslant 0\,,\label{NEC2d}\\
&&SEC_3=\frac{e^{-b}}{\kappa^2 f_T\Sigma}\left[e^{b}\left(f-f_TT+\kappa^2(p_r+2p_t+\rho)\right)\Sigma-f_{TT}T'\left(\Sigma a'+4\Sigma'\right)\right]\geqslant0\,,\label{SEC3d}\\
&&DEC_1=\frac{e^{-b}}{\kappa^2f_T\Sigma}\left[2f_{TT}T'\Sigma'-e^b\left(f-f_TT+\kappa^2(p_r-\rho)\right)\Sigma\right]\geqslant0\,,\label{DEC1d}\\
&&DEC_2=\frac{1}{2\kappa^2f_T}\left[2\kappa^2(\rho-p_t)-2f+2f_TT+\frac{e^{-b}f_{TT}T'(\Sigma a'+2\Sigma')}{\Sigma}\right]\geqslant0\,,\label{DEC2d}\\
&&DEC_3=WEC_3=\frac{1}{2\kappa^2f_T}\left[2\kappa^2\rho-f+f_TT\right]\geqslant0\,.\label{DEC3d}
\end{eqnarray}
Again here, if $f=T$, we recover the energy conditions for General Relativity, from the region inside the event horizon. 

\subsection{First Solution: black-bounce with null torsion}\label{tnula}
\par 
We will now obtain the metric functions $a(r)$, $b(r)$ and $\Sigma(r)$ so that they are consistent with the equations of motion. We want these solutions to be of the black-bounce type, so we will impose that $\Sigma(r)=\sqrt{r^2+L^2}$, as done at \cite{Visser2},  with $L \in \Re$,and furthermore, let us assume $b(r)=-\ln\left[1-\left(2m/\Sigma(r)\right)\right]$, thus leaving only the function $a(r)$ free. Note that for the diagonal tetrad case the presence of the spurious component ($\theta$, r) in equations, enables us to take  $T=0$ to get $a(r)$. Doing this we have, from the equation \eqref{te1}, 
\begin{eqnarray}
\frac{2e^{-b}\Sigma'\left(\Sigma a'+\Sigma'\right)}{\Sigma^2}=0\,,
\end{eqnarray} 
whose solution is 
\begin{eqnarray}
a(r)=a_0-\frac{1}{2}\ln\left[L^2+r^2\right]\,. \label{a1}
\end{eqnarray}
\par
Here it is important to note that, with $m>0$, taking $r\rightarrow \infty$ implies  $e^{a(r)}\rightarrow 0 $, showing that space-time for this solution is asymptotically flat, and $r\rightarrow 0$ implies $e^{a(r)}=e^{a_0}/L$, which shows that this solution is singularity-free at the origin of the coordinate system. The horizon is located in $r_H=\pm\sqrt{4m^2-L^2}$, with $0<L<2m$ or $-2m<L<0$ for $m>0$. Now, to get $\rho$, $p_r$ and $p_t$, we take, for simplification of the equations, the functions $f(r)=f_0$ and $f_T(r)=f_1$, where $f_0$ and $f_1$ are constants. Therefore, using the equations of motion \eqref{eq1}, \eqref{eq2} and \eqref{eq4},  and solution \eqref{a1}, we obtain, 
\begin{eqnarray}
&&\rho(r)=\frac{1}{2\kappa^2}\left[f_0-\frac{2f_1L^2\left(\sqrt{L^2+r^2}-4m\right)}{\left(L^2+r^2\right)^{5/2}}\right]\,,\label{rho1}\\
&&p_r(r)=-\frac{1}{2\kappa^2}\left[f_0+\frac{2f_1}{\left(L^2+r^2\right)}\right]\,,\label{pr}\\
&&p_t(r)=-\frac{1}{2\kappa^2}\left[f_0+\frac{2f_1L^2m}{\left(L^2+r^2\right)^{5/2}}+\frac{f_1\left(2L^2+r^2\right)}{2\left(L^2+r^2\right)^2}\right]\,.\label{pt}
\end{eqnarray}
Now taking the limits $r\rightarrow\infty$ we have $\rho(r)\rightarrow f_0/2\kappa^2$ and ($p_r$,$p_t$)$\rightarrow -f_0/2\kappa^2$, for $r\rightarrow 0 $ we have, 
\begin{eqnarray}
&&\rho(r)\rightarrow \frac{1}{2\kappa^2}\left(f_0-\frac{2f_1}{L^2}+\frac{8f_1m}{L^3}\right)\,,\label{rho2}\\
&&p_r(r)\rightarrow -\frac{1}{2\kappa^2}\left(f_0+\frac{2f_1}{L^2}\right)\,,\label{pr2}\\
&&p_t(r)\rightarrow \frac{1}{2\kappa^2}\left(\frac{f_1}{L^2} - \frac{2f_1m}{L^3}-f_0\right)\,.\label{pt2}
\end{eqnarray}
Here we make it explicit $\rho$, $p_r$ and $p_t$, therewith, we show that we can determine the fluid required for $T=0$. The limits $r\rightarrow 0$ and $r\rightarrow\infty$ guarantee that the solution is finite and therefore free of singularities in space-time. At future infinity, the fluid is isotropic and cosmological constant type. In origin, the density and radial and tangential pressures are constant for a given value of $L$.
\par

From equations \eqref{roeff}, \eqref{preff} and \eqref{pteff}, we can obtain the effective density and pressures for regions outside the event horizon, with $|L|<2m$, then we use \eqref{NEC1}-\eqref{DEC1} to get expressions for the energy conditions outside the event horizon, so we have,
\begin{eqnarray}
&&NEC_1=SEC_1=WEC_1=\frac{4L^2m-\sqrt{L^2+r^2}\left(2L^2+r^2\right)}{\kappa^2\left(L^2+r^2\right)^{5/2}}\geq0\,,\\
&&NEC_2=SEC_2=WEC_2=\frac{1}{4\kappa^2}\left[\frac{1}{L^2+r^2}-\frac{3L^2\left(\sqrt{L^2+r^2}-4m\right)}{\left(L^2+r^2\right)^{5/2}}\right]\geq0\,,\\
&&SEC_3=\frac{4L^2m-\sqrt{L^2+r^2}\left(2L^2+r^2\right)}{2\kappa^2\left(L^2+r^2\right)^{5/2}}\geq0\,,\\
&&DEC_1=\frac{1}{\kappa^2}\left[\frac{4L^2m}{\left(L^2+r^2\right)^{5/2}}+\frac{r^2}{\left(L^2+r^2\right)^2}\right]\geq0\,,\\
&&DEC_2=-\frac{1}{4\kappa^2}\left[\frac{1}{L^2+r^2}+\frac{5L^2\left(\sqrt{L^2+r^2}-4m\right)}{\left(L^2+r^2\right)^{5/2}}\right]\geq0\,,\\
&&DEC_3=WEC_3=-\frac{L^2\left(\sqrt{L^2+r^2}-4m\right)}{\kappa^2\left(L^2+r^2\right)^{5/2}}\geq0\,.
\end{eqnarray}
We have that, outside the event horizon $r_H$ the $NEC_1=SEC_1=WEC_1$ and $SEC_3$ are violated in $r>r_{H_+}$ and $r<r_{H_-}$. The $DEC_2$ is satisfied for $r$ near the event horizon, but still outside it, $-r_1<r<r_{H_-}$ e $r_{H_+}<r<r_1$, and is violated in $-r_1>r$ and $r>r_1$. The $NEC_2=SEC_2=WEC_2$, $DEC_1$ and $DEC_3=WEC_3$ are satisfied for $r$ external to the event horizon. There are also two cases that can be explained here: when $L=2m$, we are dealing with a wormhole with null-throat, in which case the $NEC_1=SEC_1=WEC_1$ and $SEC_3$ are violated for any value of $r$, the $NEC_2=SEC_2=WEC_2$, $DEC_1$ and $DEC_3=WEC_3$ are satisfied for all $r$, the $DEC_2$ is satisfied in $-r_1<r<r_1$ and violated in $-r_1>r$ and $r>r_1$; when $|L|>2m$ we are facing a null-throat, two-way wormhole at $r=0$, in which case the $NEC_1=SEC_1=WEC_1$, $SEC_3$, $DEC_2$ and $DEC_3=WEC_3$ are violated for any value of $r$, the $NEC_2=SEC_2=WEC_2$ and $DEC_1$ are satisfied for all $r$.

\begin{figure}[!!!h]
\centering
\begin{tabular}{rl}
\includegraphics[height=5.0cm,width=7.0cm]{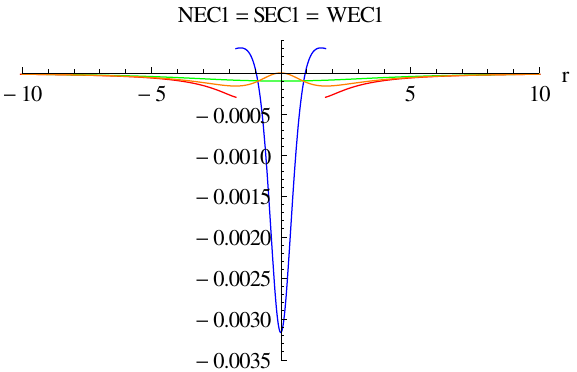}&\includegraphics[height=5.0cm,width=7.0cm]{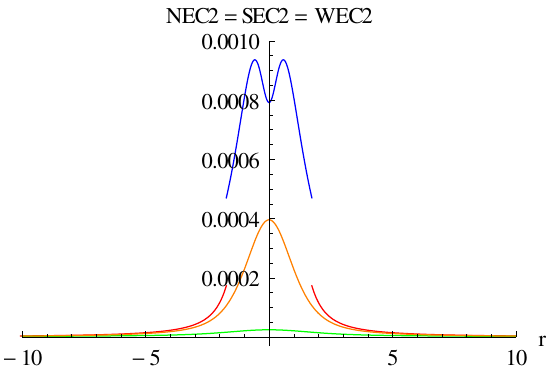}\\
\includegraphics[height=5.0cm,width=7.0cm]{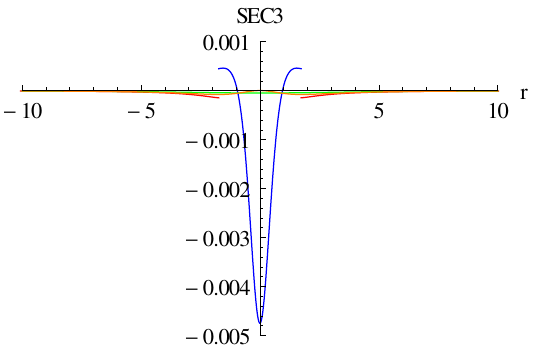}&\includegraphics[height=5.0cm,width=7.0cm]{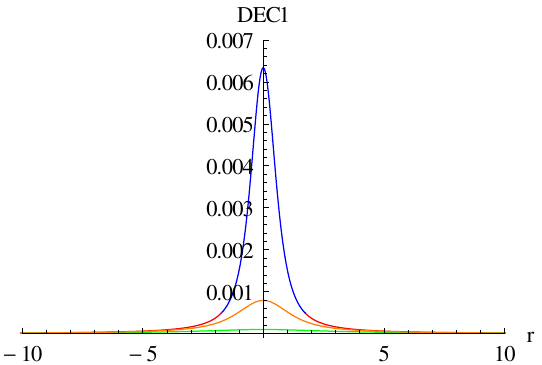}\\
\includegraphics[height=5.0cm,width=7.0cm]{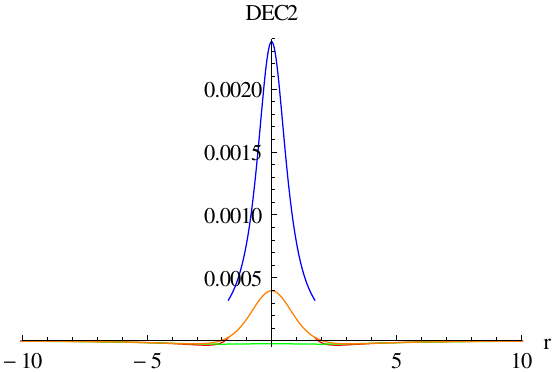}&\includegraphics[height=5.0cm,width=7.0cm]{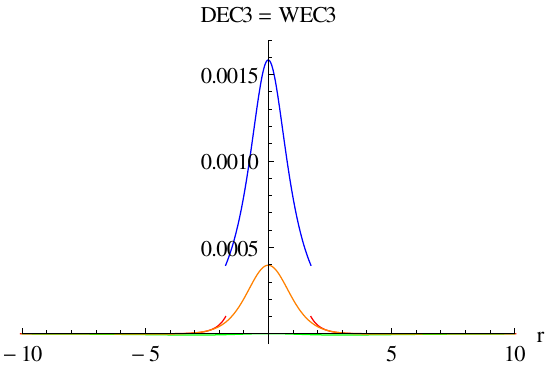}\\
\end{tabular}
\caption{\scriptsize{Graphical representation for energy conditions for this spacetime, in the region where $t$ is timelike(red) and region where $t$ is spaceliket(blue)  for $m=1$ $L=1$ and for spacetime with $L=2$(orange) and $L=4$(green).} }
\label{s1}
\end{figure}

 For regions inside $r_H$, existing only in case where $|L|<2m$, the expressions for the energy conditions are: 
\begin{eqnarray}
&&NEC_1=SEC_1=WEC_1=\frac{1}{\kappa^2}\left[\frac{1}{L^2+r^2}+L^2\left(-\frac{4m}{\left(L^2+r^2\right)^{5/2}}+\frac{1}{\left(L^2+r^2\right)^2}\right)\right]\geq0\,,\\
&&NEC_2=SEC_2=WEC_2=\frac{5r^2\sqrt{L^2+r^2}+L^2\left(6\sqrt{L^2+r^2}-4m\right)}{4\kappa^2\left(L^2+r^2\right)^{5/2}}\geq0\,,\\
&&SEC_3=\frac{3\left[r^2\sqrt{L^2+r^2}+2L^2\left(\sqrt{L^2+r^2}-2m\right)\right]}{2\kappa^2\left(L^2+r^2\right)^{5/2}}\geq0\,,\\
&&DEC_1=\frac{1}{\kappa^2}\left[\frac{4L^2m}{\left(L^2+r^2\right)^{5/2}}-\frac{r^2}{\left(L^2+r^2\right)^2}\right]\geq0\,,\\
&&DEC_2=\frac{1}{4\kappa^2}\left[\frac{3}{L^2+r^2}+L^2\left(\frac{4m}{\left(L^2+r^2\right)^{5/2}}-\frac{1}{\left(L^2+r^2\right)^2}\right)\right]\geq0\,,\\
&&DEC_3=WEC_3=\frac{1}{\kappa^2\left(L^2+r^2\right)}\geq0\,.
\end{eqnarray}
Here we have that in regions inside the event horizon, $NEC_1=SEC_1=WEC_1$ and $SEC_3$ are satisfied in a region $-r_1>r> r_{H_-}$ and $r_1<r<r_{H_+}$ and violated in $-r_1<r<r_1$. The $DEC_1$, $DEC_2$ and $DEC_3=WEC_3$ are satisfied inside the event horizon. The behavior of the energy conditions for values $m=1$ and $L=1$, with $r_H=\pm\sqrt{3}$ for the case where $|L|<2m$ are represented in red (outside the event horizon) and blue (inside the event horizon) curves in Fig.\ref{s1}. For $L=2m$ and $|L|>2m$, the energy conditions are represented by the orange($L=2$) and green($L=4$) curve, respectively, in Fig.\ref{s1}. 

\subsection{Second Solution: black-bounce with constant torsion}\label{tconstante}
\par

\begin{figure}[!!!h]
\centering
\begin{tabular}{rl}
\includegraphics[height=5cm,width=8cm]{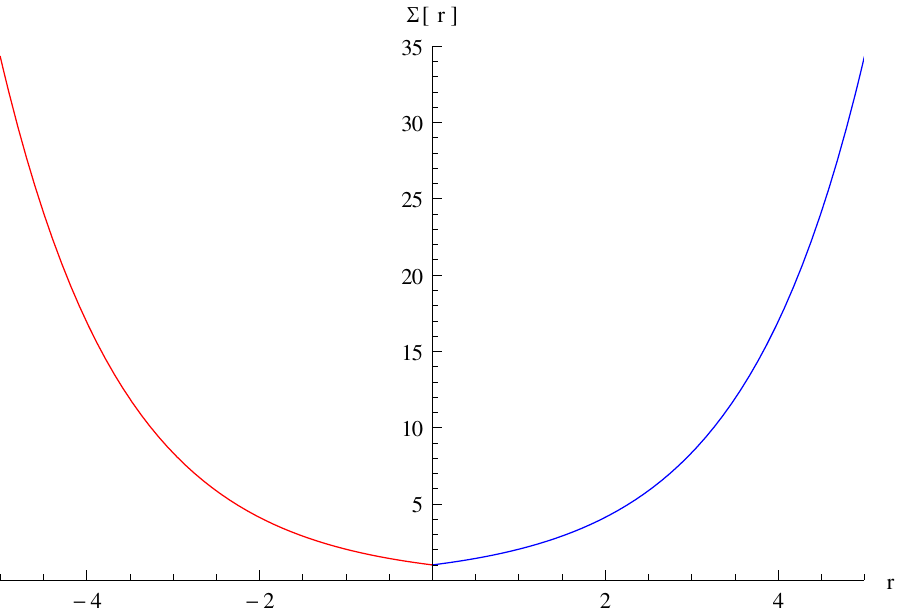}\\
\end{tabular}
\caption{\scriptsize{Graphical representation of the functions $\Sigma(r) \times r$, to $T_0=1$. In left to $r\leq0$,  in right to $r\geq0$,.} }
\label{azulvermelho}
\end{figure}

\begin{figure}[!!!h]
\centering
\begin{tabular}{rl}
\includegraphics[height=5cm,width=8cm]{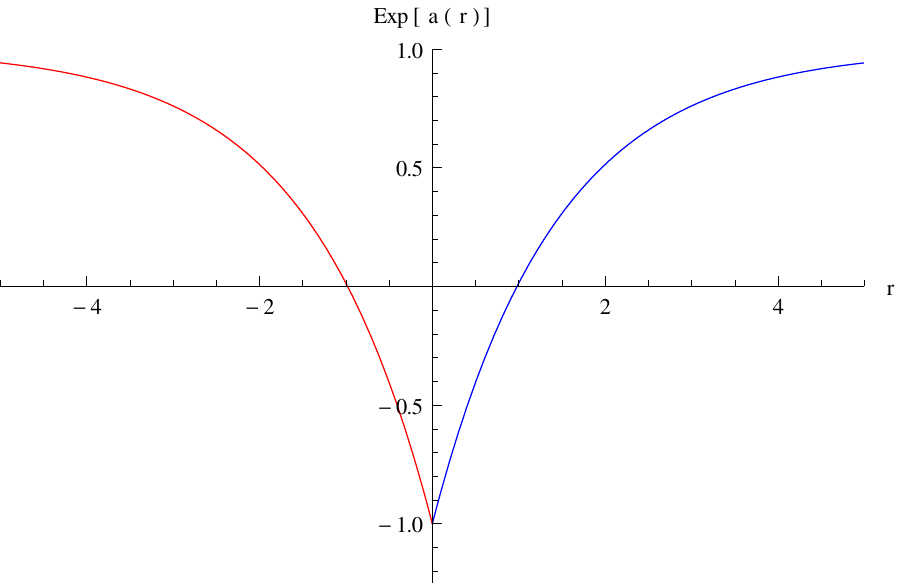}\\
\end{tabular}
\caption{\scriptsize{Graphical representation of the functions $e^{a(r)} \times r$, to $T_0=1$ and $m=1$. In left to $r\leq 0$,  in right to $r\geq 0$,.} }
\label{ab}
\end{figure}

A possible solution to the equation \eqref{te1} with constant torsion, $T=T_0$, has the form $b(r)=-\ln\left[1-\left(2m/\Sigma(r)\right)\right]$ with $\Sigma(r)=\exp\left[\pm r\sqrt{\frac{T_0}{2}}\right]$ and symmetry $a(r)=-b(r)$.
Here we have two possible situations, $r\geq 0$ and $r\leq0$. \\
For $r\geq 0$, $\exp\left[-b(r)\right]=1-2m\exp\left[-r\sqrt{\frac{T_0}{2}}\right]$. Now taking $\exp\left[-b(r)\right]=0$, the event horizon is at $r_{H_+}=\sqrt{\frac{2}{T_0}}\ln\left[2m\right]$, for $m>0$ and $T_0>0$.  
For $r\leq0$,  $\exp\left[-b(r)\right]=1-2m\exp\left[r\sqrt{\frac{T_0}{2}}\right]$. Now taking $\exp\left[-b(r)\right]=0$, the event horizon is at  $r_{H_-}=-\sqrt{\frac{2}{T_0}}\ln\left[2m\right]$, for $m>0$ and $T_0>0$.  The behavior of $\Sigma(r)=\exp\left[\pm r\sqrt{\frac{T_0}{2}}\right]$ is shown in Fig.\ref{azulvermelho}, for $r\leq0$ (red) and for $r\geq0$ (blue).
Taking the asymptotic limit of $e^{a(r)}$ we have $e^{a(r)}\rightarrow 1$ and at $r\rightarrow 0$ we have $e^{a(r)}\rightarrow 1-2m$, which shows that the solution is singularity-free throughout the space-time, as can be seen in Fig.\ref{ab}
\par
Again, we will take $f(r)=f_0$ and $f_T(r)=f_1$, solving the equation of motion \eqref{eq1}, \eqref{eq2} and \eqref{eq4}, for $r\geqslant0$ we have,
\begin{eqnarray}
&&\rho(r)=\frac{1}{2\kappa^2}\left[f_0+2f_1\left(e^{-r\sqrt{2T_0}}+2e^{-r\sqrt{\frac{T_0}{2}}}mT_0-2T_0\right)   \right]\,,\label{rho3}\\
&&p_r(r)=-\frac{1}{2\kappa^2}\left[f_0+2f_1\left(e^{-r\sqrt{2T_0}}-T_0 \right)\right]\,,\label{pr3}\\
&&p_t(r)=-\frac{1}{2\kappa^2}\left[f_0+f_1\left(e^{-r\sqrt{\frac{T_0}{2}}}m-2\right)T_0\right]. \label{pt3}
\end{eqnarray}
\par
Taking the limit $r\rightarrow 0 $ we have,
\begin{eqnarray}
\rho(r)&&\rightarrow \frac{1}{2\kappa^2}\left[f_0-2f_1\left(1-2T_0+2mT_0)\right)\right]\\
p_r(r)&&\rightarrow -\frac{1}{2\kappa^2}\left[f_0+2f_1-2f_1T_0\right]\\
p_t(r)&&\rightarrow -\frac{1}{2\kappa^2}\left[f_0+f_1(m-2)T_0\right]\,,
\end{eqnarray}
and for the limit $r\rightarrow\infty$, have
\begin{eqnarray}
\rho(r)&&\rightarrow \frac{f_0-4f_1T_0}{2\kappa^2}\,,\\
p_r(r)&&\rightarrow -\frac{f_0-2f_1T_0}{2\kappa^2}\,,\\
p_t(r)&&\rightarrow -\frac{f_0-2f_1T_0}{2\kappa^2}\,.
\end{eqnarray}
The same behavior is obtained for  $r\leq0$ on the limits $r\rightarrow-\infty$, which again shows that for $T=T_0$, the solution is singularity-free in spacetime and the fluid required for this solution is isotropic at future infinity, as well as regular at the origin of the coordinate system. \\

\begin{figure}[!!!h]
\centering
\begin{tabular}{rl}
\includegraphics[height=5.0cm,width=7.0cm]{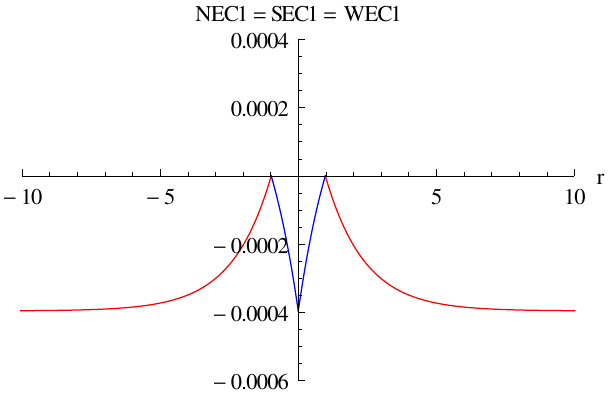}&\includegraphics[height=5.0cm,width=7.0cm]{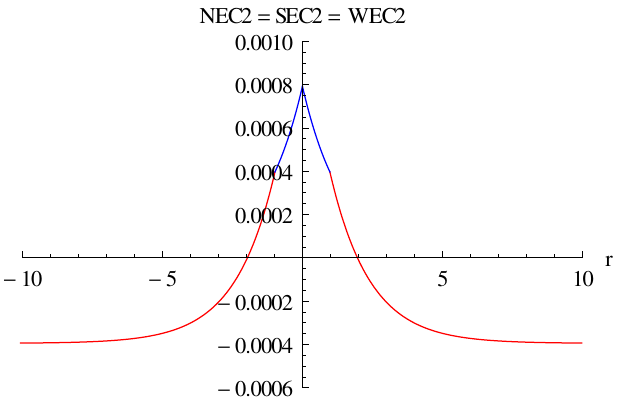}\\
\includegraphics[height=5.0cm,width=7.0cm]{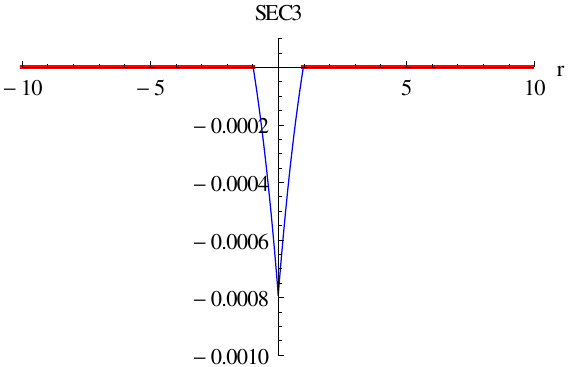}&\includegraphics[height=5.0cm,width=7.0cm]{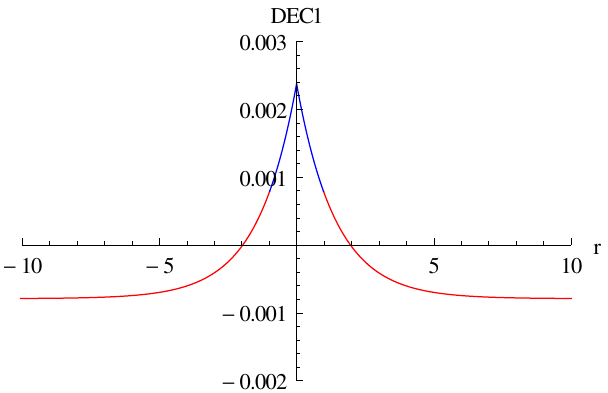}\\
\includegraphics[height=5.0cm,width=7.0cm]{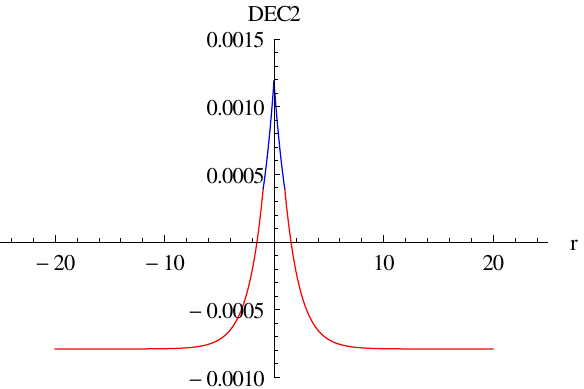}&\includegraphics[height=5.0cm,width=7.0cm]{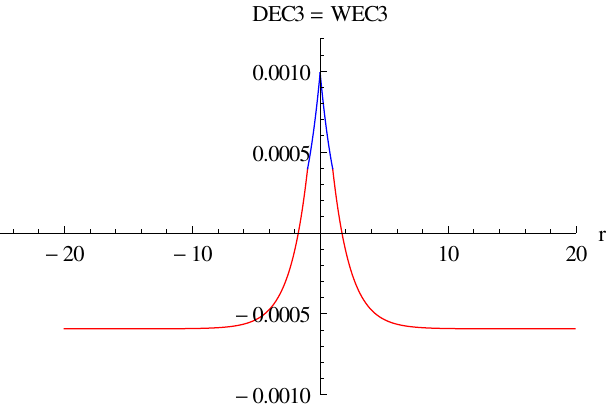}\\
\end{tabular}
\caption{\scriptsize{Graphical representation for energy conditions for this spacetime, in the region where $t$ is timelike(red) and region where $t$ is spaceliket(blue)  for $m=1$, and $T_0=1$.} }
\label{s22}
\end{figure}

Using now \eqref{roeff}, \eqref{preff} and \eqref{pteff} to calculate effective density and pressures for this solution, then using \eqref{NEC1}-\eqref{DEC3}, to calculate the energy conditions for regions outside the event horizon. There is only one sign difference here in the $r$ coordinate for regions $r\geq0$ and $r\leq0$, outside the event horizon, so the expression with subindex  ``-'' represents the energy condition for $r\geq r_{H_+}$ and the expression with subindex ``+'' denotes the energy condition for $r\leq r_{H_-}$, as follows:
\begin{eqnarray}
&&NEC_{1_\mp}=SEC_{1_\mp}=WEC_{1_\mp}=\frac{T_0}{4\kappa^2}\left[2e^{\mp r\sqrt{\frac{T_0}{2}}}m-1\right]\geq0\,,\\
&&NEC_{2_\mp}=SE_{2_\mp}=WEC_{2_\mp}=-\frac{T_0-4e^{\mp r\sqrt{\frac{T_0}{2}}}}{4\kappa^2}\geq0\,,\;\;\;\;\;SEC_{3_\mp}=0\,,\\
&&DEC_{1_\mp}=-\frac{T_0-4e^{\mp r\sqrt{\frac{T_0}{2}}}}{2\kappa^2}\geq0\,,\;\;\;DEC_{2_\mp}=\frac{e^{\mp r\sqrt{\frac{T_0}{2}}}(2+mT_0)-T_0}{2\kappa^2}\geq0\,,\\
&&DEC_{3_\mp}=WEC_{3_\mp}=\frac{2e^{\mp r\sqrt{\frac{T_0}{2}}}(4+mT_0)-3T_0}{8\kappa^2}\geq0\,.
\end{eqnarray}
The $NEC_{1_\mp}=SEC_{1_\mp}=WEC_{1_\mp}$ are violated for $r<r_{H_-}$ and $r>r_{H_+}$, whereas $NEC_{2_\mp}=SEC_{2_\mp}=WEC_{2_\mp}$, $DEC_{1_\mp}$, $DEC_{2_\mp}$ and $DEC_{3_\mp}=WEC_{3_\mp}$ are satisfied in two restricted regions $-r_1<r<r_{H_-}$ and $r_{H_+}<r<r_1$; are violated in $r<-r_1$ and $r>r_1$. The $SEC_{3_\mp}$ is satisfied everywhere $r$ outside event horizon. For regions inside event horizon the energy conditions take form:
\begin{eqnarray}
&&NEC_{1_\mp}=SEC_{1_\mp}=WEC_{1_\mp}=\frac{T_0}{4\kappa^2}\left[1-2e^{\mp r\sqrt{\frac{T_0}{2}}}m\right]\geq0\,,\\
&&NEC_{2_\mp}=SEC_{2_\mp}=WEC_{2_\mp}=-\frac{e^{\mp r\sqrt{\frac{T_0}{2}}}(mT_0-2)}{2\kappa^2}\geq0\,,\\
&&SEC_{3_\mp}=\frac{T_0}{2\kappa^2}\left[1-2e^{\mp r\sqrt{\frac{T_0}{2}}}m\right]\geq0\,,\;\;\;\;\;DEC_{1_\mp}=-\frac{T_0-4e^{\mp r\sqrt{\frac{T_0}{2}}}}{2\kappa^2}\geq0\,,\\
&&DEC_{2_\mp}=-\frac{T_0-4e^{\mp r\sqrt{\frac{T_0}{2}}}}{4\kappa^2}\geq0\,,\;\;\;DEC_{3_\mp}=WEC_{3_\mp}=\frac{e^{\mp r\sqrt{\frac{T_0}{2}}}(8-2mT_0)-T_0}{8\kappa^2}\geq0\,.
\end{eqnarray}  
Inside event horizon, in regions $r_{H_-}<r<r_{H_+}$ to $NEC_{2_\mp}$ and $SEC_{3_\mp}$ are violated. The $NEC_{1_\mp}=SEC_{1_\mp}=WEC_{1_\mp}$, $NEC_{2_\mp}=SEC_{2_\mp}=WEC_{2_\mp}$, $DEC_{1_\mp}$,$DEC_{2_\mp}$ and $DEC_{3_\mp}=WEC_{3_\mp}$ are satisfied inside event horizon.  The behavior of energy conditions for values $m=1$ and $T_0=1$, with $r_{H_\pm}=\pm\sqrt{2}\ln(2)$, can be visualized in Figs.\ref{s22}


\subsection{Teleparallel Case}
\par
Now we take the case where $f_T=1$ and $f_{TT}=0$. This is the case with Teleparallel theory.  Therefore, we require that $a(r)=-b(r)$ and for black-bounce solutions, $b(r)=-\ln\left[1-(2m/\Sigma)\right]$ and $\Sigma(r)=\sqrt{r^2+L^2}$. Therefore, the equations of motion \eqref{meq} are, 
\begin{eqnarray}
&&\frac{1}{2}\kappa^2\rho(r)=-\frac{1}{2\left(L^2+r^2\right)^{5/2}}\left[L^2\left(\sqrt{L^2+r^2}-4m\right)\right]\,,\label{m1}\\
&&\kappa^2p_r(r)=-\frac{L^2}{\left(L^2+r^2\right)^2}\,,\label{m2}\\
&&\frac{1}{2}\kappa^2p_t(r)=-\frac{L^2}{2}\left[\frac{m}{\left(L^2+r^2\right)^{5/2}}-\frac{1}{\left(L^2+r^2\right)^2}\right]\,.\label{m3}
\end{eqnarray}
Note that, for this case, $\rho=\rho^{(eff)}$, $p_r=p_r^{(eff)}$ and $p_t=p_t^{(eff)}$. The energy conditions are the same as explained in references \cite{manuelmarcos} and \cite{Visser2}, therefore,
here, NEC is violated if $\sqrt{L^2+r^2}>2m$. If $\sqrt{L^2+r^2}<2m$, then, inside possible horizons, where the coordinates $t$ and $r$ exchange their time-like with space-like characteristics, such that, $\Theta^{\;\;0}_{0}=-p_r$ and $\Theta^{\;\;1}_1=\rho$, in this region $NEC_1$ is still being violated
because it is negative everywhere, except at any possible horizon. Since the $NEC_1$ is violated, $WEC_1$, $SEC_1$ and $DEC_1$ will also be violated. For cases where $L=2m$ and $|L|>2m$, the analysis is similar to the one presented by \cite{manuelmarcos} for Simpson-Visser black-bounce case. 
We also have that scalar torsion takes form $T=2r^2/(L^2+r^2)^2$ and that $T\rightarrow 0$ in both $r\rightarrow 0$ and $r\rightarrow\infty$, which indicates regularity in spacetime.

\section{Black-Bounce solution with non-diagonal tetrad} \label{secao4}
\par
In previous section we used the dynamic field as a diagonal tetrad, this led to the appearance of a spurious component $\theta-r$ ($2-1$) in equations of motion. This is precisely due to the appropriate choice of frame reference for spherical symmetry, as seen in \cite{tamanini}. Therefore, we will consider as a good choice the tetrad matrix given in \cite{rodrigues3}, which is simply a rotation by a specific Lorentz transformation, where all equations appear consistent, which is described by
\begin{eqnarray}\label{ndtetrad}
\{e^{a}_{\;\;\mu}\}=\left[\begin{array}{cccc}
e^{a/2}&0&0&0\\
0&e^{b/2}\sin\theta\cos\phi & \Sigma\cos\theta\cos\phi &-\Sigma\sin\theta\sin\phi\\
0&e^{b/2}\sin\theta\sin\phi &
\Sigma\cos\theta\sin\phi &\Sigma\sin\theta
\cos\phi  \\
0&e^{b/2}\cos\theta &-\Sigma\sin\theta  &0
\end{array}\right]\;,
\end{eqnarray}  
where, again, by using the relationship given in (\ref{ele}), we can reconstruct the metric as
\hspace{0,2cm}
\begin{equation}
dS^2=e^{a(r)}dt^2-e^{b(r)}dr^2-{\Sigma(r)}^2\left[d\theta^{2}+\sin^{2}\left(\theta\right)d\phi^{2}\right]\label{ltbdiag}\;,
\end{equation}
where the metric parameters $\{a(r),b(r), \Sigma(r)\}$ are assumed to be functions of radial coordinate $r$ and are not time dependent. 
\par
The geometric objects established in the theory for this case are:
the non-null components of the torsional tensor  (\ref{tor}),
\begin{eqnarray}
T^{0}_{\;\;10}=\frac{a'}{2}\,,\;\;\; T^{0}_{\;\;01}=-\frac{a'}{2}\,,\;\;\;T^{2}_{\;\;21}=T^{3}_{\;\;31}=\frac{e^{b/2}-\Sigma'}{\Sigma}\,,\;\;\; T^{2}_{\;\;12}=T^{3}_{\;\;13}=\frac{\Sigma'-e^{b/2}}{\Sigma}\,.\label{tt1}
\end{eqnarray}
while the non-null components of the contorsion tensor read
\begin{eqnarray}
K_{\;\;\;\;0}^{10}=\frac{a'e^{-b}}{2}\,,\;\; K_{\;\;\;\;1}^{22}=K_{\;\;\;\;1}^{33}=\frac{e^{-b}(e^{b/2}-\Sigma')}{\Sigma}\,.\label{ctt1}
\end{eqnarray}
The non-null components of the tensor $S_{\alpha}^{\;\;\mu\nu}$:
\begin{eqnarray}
S_{0}^{\;\;01}=\frac{e^{-b}(e^{b/2}-\Sigma')}{\Sigma}\,,\;\;S_{2}^{\;\;12}=S_{3}^{\;\;13}=\frac{e^{-b}\left(a'\Sigma-2e^{b
2}+2\Sigma'\right)}{4\Sigma}\,.\label{st1}
\end{eqnarray}
\par
From the definition of the torsion scalar (\ref{t1}), one gets
\begin{equation}
T=\frac{2e^{-b}}{\Sigma^2}\left[\left(e^{b/2}-\Sigma'\right)\left(e^{b/2}-\Sigma a'-\Sigma'\right)\right] \label{te}\,.
\end{equation}
We note here that in general, the scalar torsion is an arbitrary function of the radial coordinate $r$.
\par
The equation of motion \eqref{meq}, with $\Theta_{\nu}^{\;\;\mu}=diag\left[\rho(r), -p_r(r), -p_t(r), -p_t(r)\right]$,  are now: 
\begin{eqnarray}
&&\frac{e^{-b}}{2\Sigma^2}\Big[2\Sigma f_{TT}T'\left(e^{b/2}-\Sigma'\right)+f_T\Big(2e^{b/2}\Sigma'-2(\Sigma')^2+\Sigma\left(e^{b/2}a'+(b'-a')\Sigma'-2\Sigma''\right)\Big)\Big]\nonumber\\
&&+\frac{f}{4}=\frac{1}{2}\kappa^2\rho\,,\label{eqnd1}\\
&&\frac{e^{-b}f_T}{2\Sigma^2}\left[-2\Sigma'\left(\Sigma a'+\Sigma'\right)+e^{b/2}\left(\Sigma a'+2\Sigma'\right)\right]+\frac{f}{4}=-\frac{1}{2}\kappa^2p_r\,,\label{eqnd2}\\
&&\frac{e^{-b}}{8\Sigma^2}\Big[-2\Sigma f_{TT}T\left(-2e^{b/2}+\Sigma a'+2\Sigma'\right)-f_T\Big(\left(4e^{b/2}-4\Sigma'\right)^2+\Sigma^2\left((a')^2-a'b'+2a''\right)+\nonumber\\
&&\Sigma\left(4\Sigma''-2b'\Sigma'+a'\left(6\Sigma'-4e^{b/2}\right)\right)\Big]+\frac{f}{4}=-\frac{1}{2}\kappa^2 p_t\,.\label{eqnd3}
\end{eqnarray}

\par

For a first approach to where tetrads are non-diagonal, lets assume that $a(r)=-b(r)$.  Remembering that $f_{TT}$ and $f_T$ are related as follows, $f_{TT}=\left(\frac{\partial T}{\partial r}\right)^{-1}\frac{\partial f_T}{\partial r}$.  Subtracting \eqref{eqnd1} from \eqref{eqnd2} we have:
\begin{eqnarray}
\frac{e^{-b}}{\Sigma}\left[f_{T}'\left(-e^{b/2}+\Sigma'\right)+f_T\Sigma''\right]=0\,,
\end{eqnarray}
solving for $f_T$ we have 
\begin{eqnarray}
f_T(r)=\exp\left[\int\frac{dr \Sigma''}{e^{\frac{1}{2}b}-\Sigma'}\right] \,. \label{fTND}
\end{eqnarray}
Here we see that there is now a possibility in which $g_{00}(r)=e^{a(r)}=-g^{11}(r)=e^{-b(r)}$, that is, $a(r)=-b(r)$, and $f(T)$ is still non-linear in $T$, violating the no-go theorem established in \cite{ednaldo}. This is clearly possible due to the new expression for function that determines the area related to the metric, $\Sigma(r)$, where the particular case where $\Sigma(r)=r$, and the area in this case is $4\pi r^2$, \eqref{fTND} provides $f_T=1$, back to the linear case $f(T)=T$. In general, $\Sigma(r)\neq r$, then $f(T)$ is not linear anymore. This is a new result in this work.
\par
\subsection*{Case for  $-\sqrt{4m^2-L^2}\geq r$  and  $ r\geq\sqrt{4m^2-L^2}$ :}\label{casofora}
\par 
For black-bounce space-time we have, $\Sigma(r)=\sqrt{r^2+L^2}$ with $b(r)=-\ln\left[1-\left(2m/\Sigma(r)\right)\right]$, from which we get the scalar torsion $T(r)$,
\begin{eqnarray}
&&T(r)=-\frac{2}{(L^2+r^2)^2\left(\sqrt{L^2+r^2}-2m\right)}\left(\sqrt{L^2+r^2}-r\sqrt{1-\frac{2m}{\sqrt{L^2+r^2}}}\right)\Bigg(-L^2-r^2\nonumber\\
&&+2m\sqrt{L^2+r^2}+r\sqrt{L^2+r^2}\sqrt{1-\frac{2m}{\sqrt{L^2+r^2}}}\Bigg)\,, \label{TescND}
\end{eqnarray}
with,  $-\sqrt{4m^2-L^2}\geq r$ e $ r\geq\sqrt{4m^2-L^2}$ and $-2m \leq L\leq 2m$. Taking the limit where $r\rightarrow\infty$ we have $T=0$ and for $r\rightarrow 0$ we have $T=2/L^2$, which characterizes the solution to be asymptotically flat and regular throughout the space-time.
\par
\begin{figure}[!!!h]
\centering
\begin{tabular}{rl}
\includegraphics[height=5cm,width=7cm]{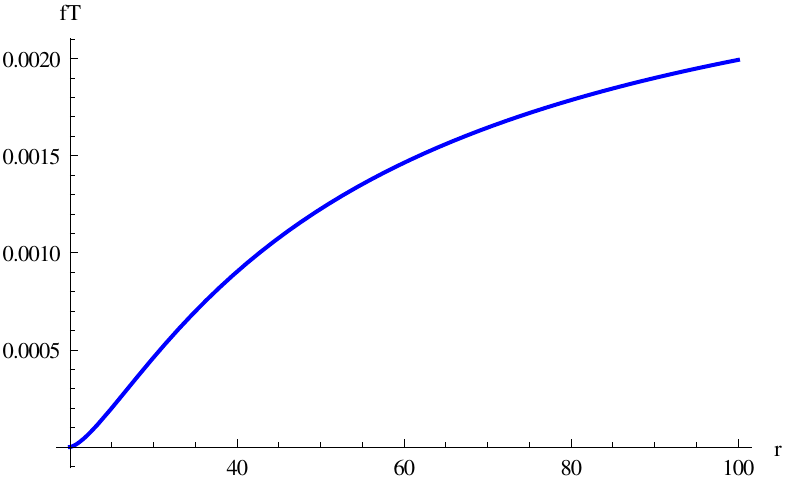}&\includegraphics[height=5cm,width=7cm]{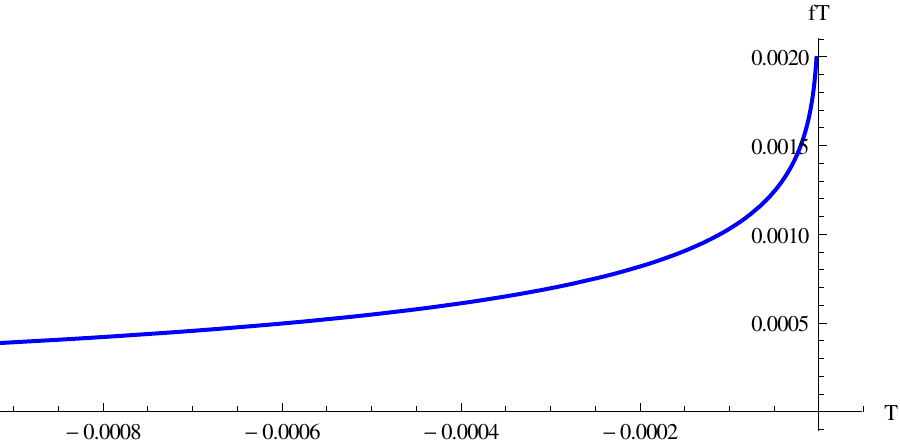}\\
\end{tabular}
\caption{\scriptsize{Graphical representation of the functions $f_T \times r$ and $f_T \times T$, to $m=10$ and $L=1$.} }
\label{fig1}
\end{figure}

 
We numerically calculate the function \eqref{fTND}, using $b(r)$, and then the chain rule $df/dT=(df/dr)(dr/dT)$, integrating $\int f_T(dT/dr)dr$ to determine $f(r)$. The form of $f_T$ is shown in Fig.\ref{fig1} for $f_T \times r$ and $f_T \times T$.
\par
\begin{figure}[!!!h]
\centering
\begin{tabular}{rl}
\includegraphics[height=5cm,width=7cm]{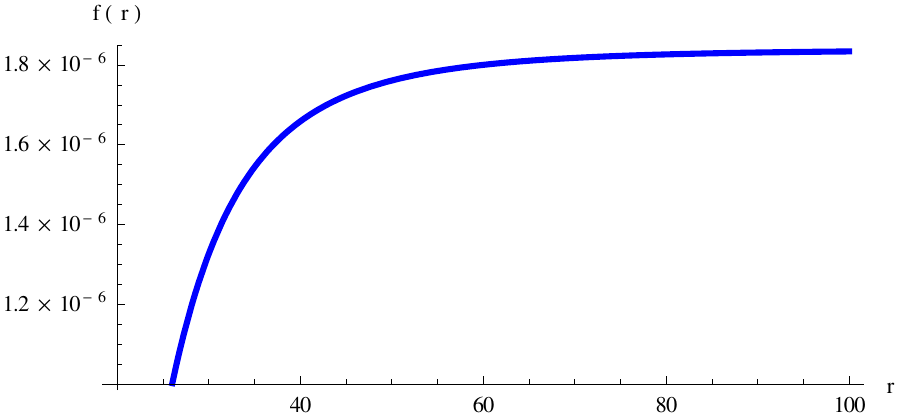}&\includegraphics[height=5cm,width=7cm]{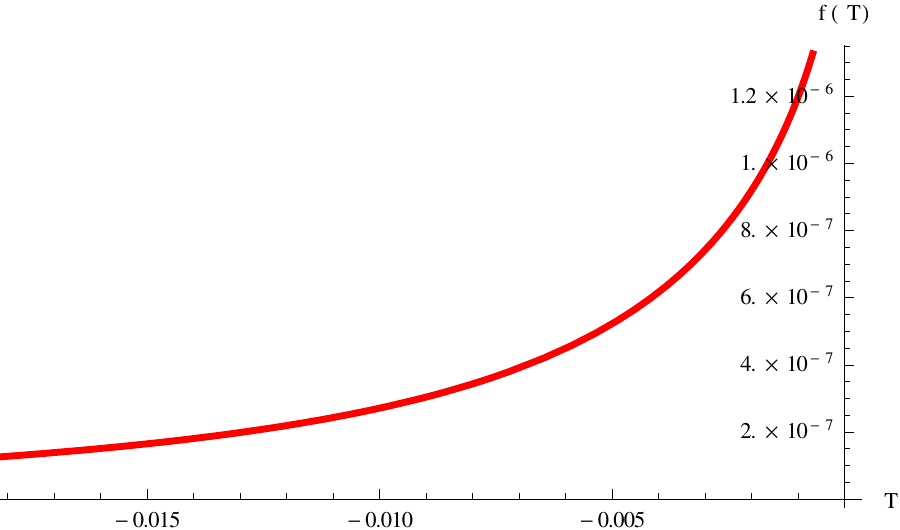}\\
\end{tabular}
\caption{\scriptsize{Graphical representation of the functions $f(r) \times r$ (left) and $f(T) \times T$ (right), to $m=10$ and $L=1$.} }
\label{figfxT}
\end{figure}

The form of $f(r)$ is shown in Fig.\ref{figfxT} which we clearly see, through the parametric plot of $f(T)\times T$ (in red), that we are dealing with a nonlinear theory in $T$ and therefore, we are not dealing with Teleparallel Theory.
\par
\subsection*{Case for $-\sqrt{4m^2-L^2}\leq r \leq\sqrt{4m^2-L^2}$ :}\label{casodentro}
\par 
We will use again $\Sigma(r)=\sqrt{r^2+L^2}$ so we have a black-bounce type solution, however, now with, $b(r)=-\ln\left[-1+\left(2m/\Sigma(r)\right)\right]$, the scalar torsion $T(r)$ takes the form,
\begin{eqnarray}
&&T(r)=-\frac{2}{(L^2+r^2)^2\left(-2m+\sqrt{L^2+r^2}\right)}\left(\sqrt{L^2+r^2}-r\sqrt{-1+\frac{2m}{\sqrt{L^2+r^2}}}\right)\Bigg(-L^2-r^2\nonumber\\
&&+2m\sqrt{L^2+r^2}+r\sqrt{L^2+r^2}\sqrt{-1+\frac{2m}{\sqrt{L^2+r^2}}}\Bigg)\,, \label{TescND2}
\end{eqnarray}
where in this case,  $-\sqrt{4m^2-L^2}\leq$ r $\leq\sqrt{4m^2-L^2}$ for $-2m<L<2m$. In addition, $r=0$ for  $L=\pm2m$ and $-2m \leq r\leq 2m$ for $L=0$.At limit $r\rightarrow 0$ we have $T=2/L^2$ and for $r\rightarrow\infty$, $T=0$, which shows that the scalar torsion is regular throughout spacetime, also indicating a regular solution. 
\par

\begin{figure}[!!!h]
\centering
\begin{tabular}{rl}
\includegraphics[height=5cm,width=7cm]{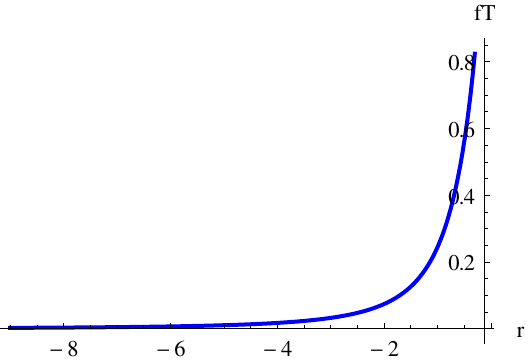}&\includegraphics[height=5cm,width=7cm]{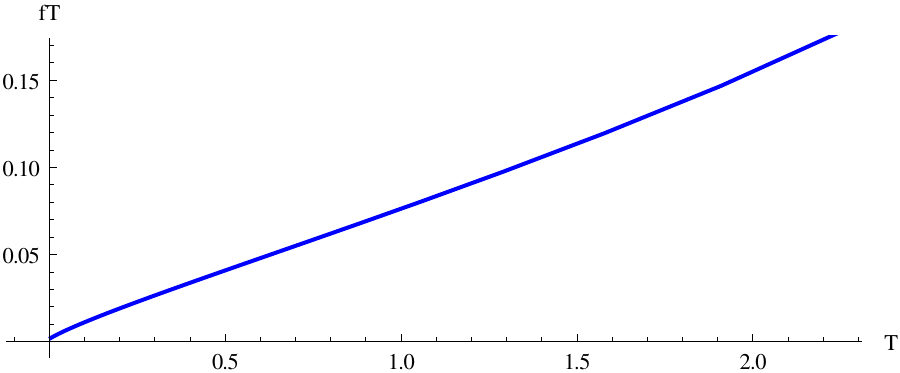}\\
\end{tabular}
\caption{\scriptsize{Graphical representation of the functions $f_T \times r$ and $f_T \times T$, to $m=10$ and $L=1$.} }
\label{fig2}
\end{figure}

\begin{figure}[!!!h]
\centering
\begin{tabular}{rl}
\includegraphics[height=5cm,width=7cm]{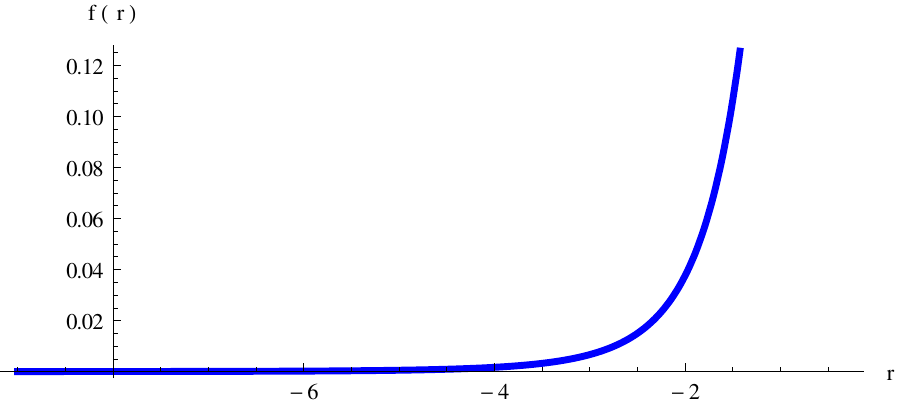}&\includegraphics[height=5cm,width=7cm]{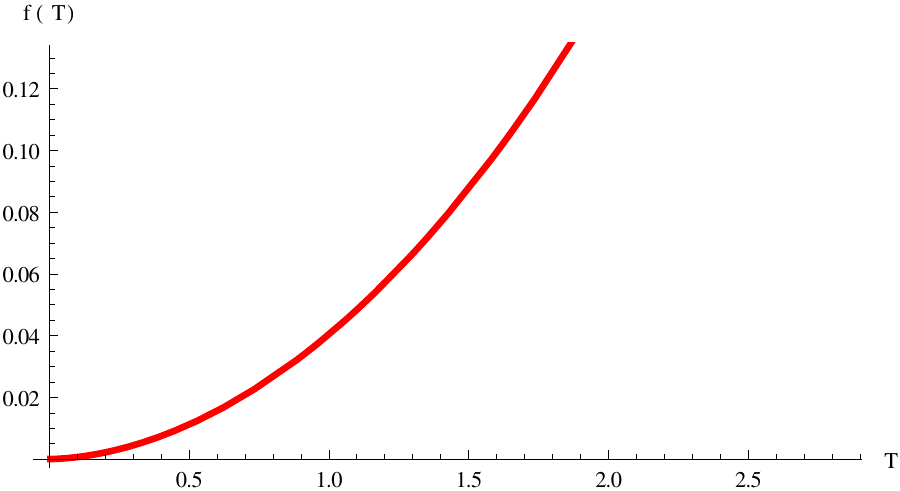}\\
\end{tabular}
\caption{\scriptsize{Graphical representation of the functions $f(r) \times r$ and $f(T) \times T$, to $m=10$ and $L=1$.} }
\label{fig3}
\end{figure}

Subtracting \eqref{eqnd2} from \eqref{eqnd1} we can obtain $f_T(r)$, therefore, 
\begin{eqnarray}
&&f_T(r)=\exp\Bigg\{-\int\frac{L^2\left(\sqrt{L^2+r^2}-2m\right)}{L^2+r^2}\Bigg[L^2 \sqrt{\frac{2m}{\sqrt{L^2+r^2}}-1} \nonumber\\
&&+r\Bigg(\sqrt{L^2+r^2}-2m+r\sqrt{\frac{2m}{\sqrt{L^2+r^2}}-1}\Bigg)\Bigg]^{-1}dr\Bigg\}\,.\label{fTND2}
\end{eqnarray}
We plot the behavior of \eqref{fTND2} in Fig.\ref{fig2} for $f_T \times r$ and $f_T \times T$, which was calculated numerically using $b(r)$.  Then, using the chain rule $df/dT=(df/dr)(dr/dT)$, integrating $\int f_T(dT/dr)dr$  we finally determine $f(r)$ and plot its behavior for $f(r) \times r$ and $f(T) \times T$ in Fig.\eqref{fig3}, which clearly shows that we are facing a non-linear theory in $T$.  
The energy conditions for the two cases presented here, with $r$ outside the event horizon (-$\sqrt{4m^2-L^2}\geq r$  and  $ r\geq\sqrt{4m^2-L^2}$) and inside event horizon ($-\sqrt{4m^2-L^2}\leq r \leq\sqrt{4m^2-L^2}$) are the same as shown in \cite{Visser2} and \cite{manuelmarcos} for GR. The region inside event horizon requires that $|L|<2m$. Thus, we have that $NEC_1$ is violated for all $r$ inside and outside the event horizon. The $NEC_2$, $SEC_3$ and $DEC_1$ are satisfied for all $r$ inside and outside the event horizon. The $DEC_2$ and $DEC_3$ are violated for $|r|>>r_H$. To a null-throat wormhole, where $L=2m$, the $NEC_1$ is violated for all values of $r$, the $DEC_2$ and $DEC_3$ are satisfied for $-r_1<r<r_1$ and violated to $-r_1>r$ and $r>r_1$. To $|L|>2m$, wormhole with no null throat, $NEC_1$, $DEC_2$ and $DEC_3$ are violated, $NEC_2$, $SEC_3$ and $DEC_1$ are satisfied, for any value of $r$. 
\subsection*{Quadratic case in $T$:}
\par
We impose a quadratic model on $T$ to check the behavior of a possible fluid for theory. Making  $f(T)=T+a_0T^2$, where $a_0$ is constant, $a(r)=-b(r)$, with $b(r)=-\ln\left[\left(1-2m/\Sigma\right)^2\right]$ and $\Sigma(r)=\sqrt{r^2+L^2}$. Using \eqref{te} we obtain the scalar torsion,
\begin{eqnarray}
T(r)&=&\frac{2\left(1-\frac{2m}{\sqrt{L^2+r^2}}\right)^2}{L^2+r^2}\left[\frac{1}{1-\frac{2m}{\sqrt{L^2+r^2}}}-\frac{r}{\sqrt{L^2+r^2}}\right]\Bigg[\frac{1}{1-\frac{2m}{\sqrt{L^2+r^2}}}-\frac{r}{\sqrt{L^2+r^2}}\nonumber\\
&&-\frac{4mr}{L^2+r^2-2m\sqrt{L^2+r^2}}\Bigg]\,,
\end{eqnarray}
and taking the limits, $r\rightarrow0$ and $r\rightarrow\infty$ we have $T(r)=2/L^2$ and $T(r)=0$ respectively, which indicates regularity throughout space-time. From $e^{-b(r)}=0$ we get the horizons for this solution, which are in $r_H=\pm\sqrt{4m^2-L^2}$ com $-2m<L<2m$. Now, solving the eqs. \eqref{eqnd1}, \eqref{eqnd2} and \eqref{eqnd3} we obtain density $\rho(r)$ and pressures $p_r(r)$ and $p_t(r)$ and calculate the limits  $r\rightarrow0$ getting, 
\begin{eqnarray}
\rho && \rightarrow \frac{8L\left(4a_0+L^2\right)m-8\left(4a_0+L^2\right)m^2-L^2\left(6a_0+L^2\right)}{L^6\kappa^2}\,,\\
p_r&&\rightarrow -\frac{2a_0+L^2}{L^2\kappa^2}\,,\\
p_t&& \rightarrow \frac{(6a_0-2Lm+L^2)L-8a_0m}{L^5\kappa^2}\,,
\end{eqnarray}  
and in the limit for $r\rightarrow\infty$, both $\rho$ and $p_r$ and $p_t$ are null, which indicates that it is a well behaved isotropic fluid. 
\par

The energy conditions for this solution are obtained from the equations \eqref{roeff}, \eqref{preff} and \eqref{pteff}, then, using \eqref{NEC1}-\eqref{DEC3}. For regions outside the event horizon the expressions are:
\begin{eqnarray}
&&NEC_1=-\frac{2L^2}{\kappa^2\left(L^2+r^2\right)^{7/2}}\left[(4m^2+r^2)\sqrt{L^2+r^2}+L^2\left(\sqrt{L^2+r^2}-4m\right)-4mr^2\right],\\
&&NEC_2= \frac{2m}{\kappa^2\left(L^2+r^2\right)^{7/2}}\left[3L^4+4mr^2\sqrt{L^2+r^2}+L^2\left(3r^2-4m\sqrt{L^2+r^2}\right)\right]\,,\\
&&SEC_3=\frac{4m}{\kappa^2\left(L^2+r^2\right)^{7/2}}\left[L^4+2mr^2\sqrt{L^2+r^2}+L^2\left(r^2-2m\sqrt{L^2+r^2}\right)\right]\,,\\
&&DEC_1=\frac{8m}{\kappa^2\left(L^2+r^2\right)^{7/2}}\left[L^4+mr^2\sqrt{L^2+r^2}+L^2\left(r^2-m\sqrt{L^2+r^2}\right)\right]\,,\\
&&DEC_2=-\frac{2L^2}{\kappa^2\left(L^2+r^2\right)^{7/2}}\left[(4m^2+r^2)\sqrt{L^2+r^2}+L^2\left(\sqrt{L^2+r^2}-5m\right)-5mr^2\right]\,,\\
&&DEC_3=\frac{1}{\kappa^2\left(L^2+r^2\right)^{7/2}}\left[8L^2m\left(L^2+r^2\right)-\sqrt{L^2+r^2}\left(L^4-4m^2r^2+L^2\left(8m+r^2\right)\right)\right]
\end{eqnarray}
With $NEC_{1,2}=WEC_{1,2}=SEC_{1,2}$ and $DEC_3=WEC_3$. For a regular black hole where $|L|<2m$ with event horizon in $r_{H_{\pm}}$, the $NEC_1$ is violated outside the event horizon and satisfied in $r_{H_-}<r<r_{H_+}$, here $NEC_1$ has finite value at $r=0$. The $NEC_2$, $SEC_3$, $DEC_1$ and $DEC_3=WEC_3$ sare satisfied for all values of $r$ inside and outside the event horizon. The $DEC_2$ is satisfied for an outside region close to the event horizon $-r_1<r<r_{H_-}$ and $r_{H_+}<r<r_1$ and violated to $-r_1>r$ and $r>r_1$. To a null-throat wormhole, $L=2m$, the $NEC_1$ is violated for any value of the coordinate $r$ and $NEC_2$, $SEC_3$, $DEC_1$ and $DEC_3=WEC_3$ are satisfied for any value of $r$ however, the $DEC_2$ is satisfied only in a region bounded by $-r_1<r<r_1$ and violated for values $-r_1>r$ and $r>r_1$. For a two-way wormhole at $r=0$, with no null throat, $|L|>2m$, the $NEC_1$ and $DEC_2$ are violated for every value of the radial coordinate and $NEC_2$, $SEC_3$, $DEC_1$ and $DEC_3=WEC_3$ are satisfied for all $r$. In regions inside the event horizon, valid only for $|L|<2m$ the energy conditions are:
\begin{eqnarray}
&&NEC_1=\frac{2L^2}{\kappa^2\left(L^2+r^2\right)^{7/2}}\left[(4m^2+r^2)\sqrt{L^2+r^2}+L^2\left(\sqrt{L^2+r^2}-4m\right)-4mr^2\right]\,,\\
&&NEC_2=\frac{2}{\kappa^2\left(L^2+r^2\right)^{7/2}}\left[4m^2r^2\sqrt{L^2+r^2}+L^2(L^2+r^2)\left(\sqrt{L^2+r^2}-m\right)\right]\,,\\
&&SEC_3=\frac{4}{\kappa^2\left(L^2+r^2\right)^{5/2}}\left[2m^2\sqrt{L^2+r^2}+L^2\left(\sqrt{L^2+r^2}-3m\right)\right]\,,\\
&&DEC_1=\frac{8m}{\kappa^2\left(L^2+r^2\right)^{7/2}}\left[L^4+mr^2\sqrt{L^2+r^2}+L^2\left(r^2-m\sqrt{L^2+r^2}\right)\right]\,,\\
&&DEC_2=\frac{2L^2m}{\kappa^2\left(L^2+r^2\right)^{5/2}}\,,\;\;\;\;DEC_3=WEC_3=\frac{L^4+\left(L^2+4m^2\right)r^2}{\kappa^2\left(L^2+r^2\right)^3}\,.
\end{eqnarray}
The $NEC_1$ is satisfied in $r_{H_-}<r<r_{H_+}$ and is finite in $r=0$. The $NEC_2$, $SEC_3$, $DEC_1$, $DEC_2$ and $DEC_3=WEC_3$ are satisfied for all values $r_{H_-}<r<r_{H_+}$. 

\begin{figure}[!!!h]
\centering
\begin{tabular}{rl}
\includegraphics[height=5.0cm,width=7.0cm]{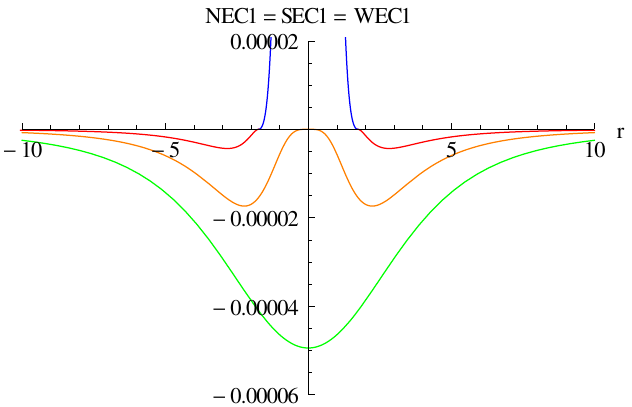}&\includegraphics[height=5.0cm,width=7.0cm]{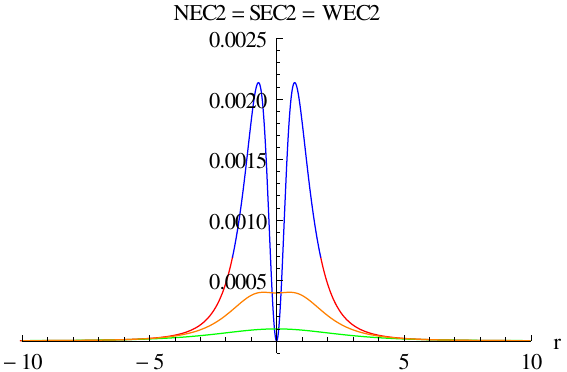}\\
\includegraphics[height=5.0cm,width=7.0cm]{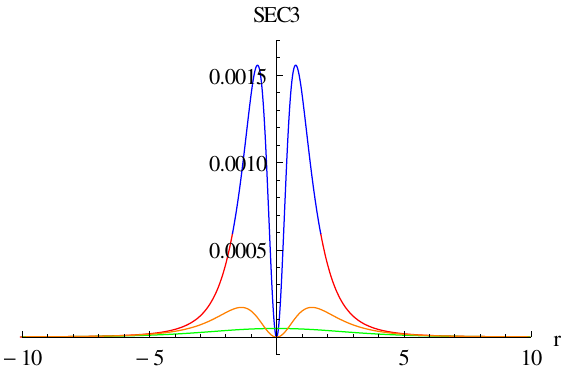}&\includegraphics[height=5.0cm,width=7.0cm]{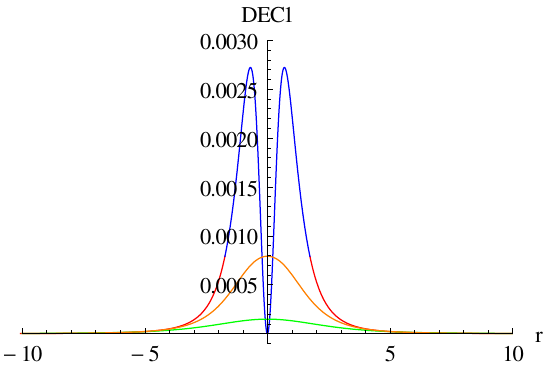}\\
\includegraphics[height=5.0cm,width=7.0cm]{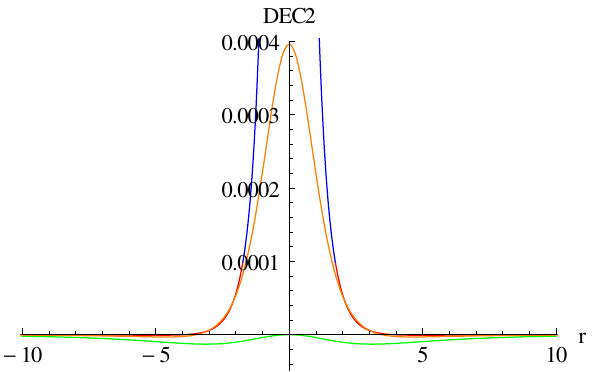}&\includegraphics[height=5.0cm,width=7.0cm]{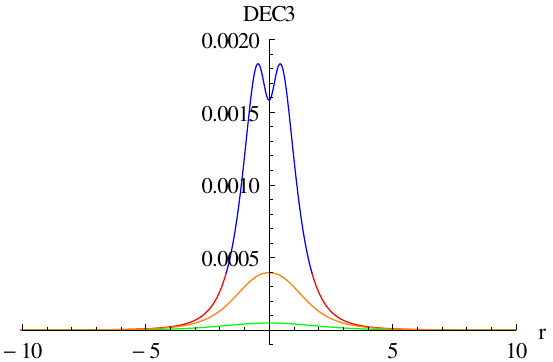}\\
\end{tabular}
\caption{\scriptsize{Graphical representation for energy conditions for this spacetime, in the region where $t$ is timelike(red) and region where $t$ is spaceliket(blue)  for $m=1$ $L=1$ and for spacetime with $L=2$(orange) and $L=4$(green).} }
\label{s4}
\end{figure}

The behavior of the energy conditions for values $m=1$ and $L=1$, with $r_H=\pm\sqrt{3}$ for the case where $|L|<2m$ are represented in the red (outside the event horizon) and blue (inside the event horizon) curves in Fig.\ref{s4}. For $L=2m$ and $|L|>2m$, the energy conditions are represented by the orange($L=2$) and green($L=4$) curve, respectively, in Fig.\ref{s4}. 

\section{Conclusion}
\par

We explore in this paper the called black-bounce solutions in the context of the theory of $f(T)$ gravity in two different frames of reference, the Sec.\ref{secao3} with diagonal tetrads and Sec.\ref{secao4} with non-diagonal tetrads, both frames describe spherically symmetric and static metrics in spherical coordinates. 
\par 
The first solution presented in Sec.\ref{secao3} is a case with null torsion in the context of Simpson-Visser black-bounce with $b(r)=-\ln\left[1-\left(2m/\Sigma\right)\right]$ and $\Sigma(r)=\sqrt{r^2+L^2}$, where $a(r)$ is a free function that was obtained later. We calculate the density and the radial and tangential pressures, showing that the fluid required for this solution is well behaved and singularity free in space-time, at future infinity it is isotropic and cosmological constant type, at origin the density and pressures are constant and governed by a given value of the parameter $L$. The energy conditions were obtained from the effective density and pressures and analyzed for the case of a regular black hole with $|L|<2m$. In this case $NEC_1$ and $SEC_1$ are violated outside the event horizon $r_H$ and satisfied only in a small region inside the horizon. It is known that $SEC_3$ is violated inside event horizon for regular black hole solutions in GR, while $WEC$ is, in some cases, violated throughout space-time, but this is not a rule for black-bounce type space-time. For null-throat wormholes, the $NEC_1$ and $SEC_3$ are violated for all $r$. For a two-way wormhole, with $|L|>2m$, the $NEC_1$, $SEC_3$, $DEC_2$ and $DEC_3$ are violated for all values of $r$. We plotted graphs in Fig.\ref{s1} for energy conditions of this solution for the three cases specified here. 
\par
As second solution for diagonal tetrad, we present a specific type of black-bounce where $\Sigma(r)=\exp\left[\pm r\sqrt{T_0/2}\right]$ for this solution, two cases are possible, $r\geqslant 0$ and $r\leqslant 0$ with horizons in $r_{H_{\pm}}=\pm\sqrt{2/T_0}\ln\left[2m\right]$ respectively. For both cases this solution is regular throughout the space-time and the associated fluid is singularity-free and isotropic. The energy conditions were analyzed according to density and effective pressures and for both cases the $NEC_1$ is violated outside event horizon while the $NEC_2$, $DEC_1$, $DEC_2$ and $DEC_3$ are violated for $r$ far away from the event horizon. The $NEC_2$ and $SEC_3$ are violated inside event horizon, as in the case of GR. The case for Teleparallel theory was also addressed, being analogous to GR.  
\par 
In Sec.\ref{secao4} the equations of motion of $f(T)$ were written for non-diagonal tetrads. Two cases for Simpson-Visser black-bounce, with $a(r)=-b(r)$, were explored. The first, external to event horizon, with  $-\sqrt{4m^2-L^2}\geq r$  and  $ r\geq\sqrt{4m^2-L^2}$, the scalar torsion vanishes at infinity and at origin is finite regulated by the parameter $L$, characterizing this solution as regular throughout the space-time. We numerically calculate the function $f(T)$ and plot in Fig.\ref{fig3}, making explicit its non-linear character, as expected for $f(T)$ gravity. The second case, internal to the event horizon, again the limits for scalar torsion have been obtained, the space-time is asymptotically flat and regular at the origin of coordinate system, governed by $L$. We plot $f(T)$ making its non-linearity explicit again. The energy conditions for the two solutions are also analogous to those obtained in GR for black-bounce space-time analyzed in \cite{Visser2} and \cite{manuelmarcos}. We also study a quadratic model for $f(T)$ with symmetry $a(r)=-b(r)$ and $b(r)=-\ln\left[\left(1-2m/\Sigma\right)^2\right]$, defined in reals. The event horizon is at $r_H=\pm\sqrt{4m^2-L^2}$. The solution is asymptotically flat, as we saw by taking limits on $T(r)$, and finite at the origin, parametrized by $L$. The characteristic fluid of this solution is isotropic and well behaved. For a regular black hole where $|L|<2m$ the $NEC_1$ is violated only outside $r_H$ and $DEC_2$ for $r$ far away from $r_H$. The $NEC_1$ is violated for the null-throat wormhole case. To $|L|>2m$, the $NEC_1$ and $DEC_2$ are violated in all space-time. Another interesting result in this section is that due to the possibility that the area associated with the metric is different from $4\pi r^2$, the no-go theorem established on the usual $f(T)$ is violated, and the new possibility $g_{00}=-g^{11}$,for the metric components. 
\par
We present and explore several black-bounce solutions for theory $f(T)$ gravity. It is important to note that $SEC_3$ is at least partially satisfied inside event horizon for the solutions presented here, except for the solution with constant torsion. This energy condition is always violated for regular black hole solutions inside event horizon so, as we have seen, the $SEC_3$ violation is a particular case for black-bounce.

\vspace{1cm}

{\bf Acknowledgement}:  M. E. R. thanks CNPq for partial financial support.


%


\begin{thebibliography}{99}

\bibitem{Sch}
K. Schwarzschild, Sitzungsber. K. Preuß. Akad,{\it{ Über das Gravitationsfeld eines Massenpunktes nach der Einsteinschen Theorie}}, Berlin (Math. Phys.) 1916, {\bf 189} (1916).  [\href{https://arxiv.org/abs/physics/9905030}
{{\tt 	arXiv:physics/9905030 }}].
\bibitem{Reissner}
H. Reissner, {\it{Über die Eigengravitation des elektrischen Feldes nach der Einsteinschen Theorie}} Ann. Physik, {\bf 50}, 106-120 (1916), [\href{https://onlinelibrary.wiley.com/doi/abs/10.1002/andp.19163550905}{{\tt andp.19163550905}}].
\bibitem{Nordstrom}
G. Nordstrom, {\it{On the Energy of the Gravitation field in Einstein's Theory}},  Proc. Kon. Ned. Akad. Wet., {\bf 20}, (1918) 1238-1245, [\href{http://adsabs.harvard.edu/abs/1918KNAB...20.1238N}{{\tt 1918KNAB...20.1238N}}].

\bibitem{Kerr}
R. P. Kerr,{\it{Gravitational field of a spinning mass as an example of algebraically special metrics  }}, Phys. Rev. Lett. {\bf 11}, 237 (1963), [\href{https://journals.aps.org/prl/abstract/10.1103/PhysRevLett.11.237}{{\tt DOI: 10.1103/PhysRevLett.11.237}}].

\bibitem{Horizon}
Event Horizon Telescope Collaboration,{\it{First M87 Event Horizon Telescope Results. I. The Shadow of the Supermassive Black Hole}}, Astrophys.J. {\bf 1} 875 (2019), L1, [\href{https://arxiv.org/abs/1906.11238}{{\tt arXiv:1906.11238 }}].

\bibitem{Sakharov}
AD Sakharov, {\it{The initial stage of an expanding Universe and the appearance of a nonuniform distribution of matter}}, Sov. Phys. JETP 22, {\bf 241} (1966), [\href{http://jetp.ras.ru/cgi-bin/dn/e_022_01_0241.pdf}{{\tt http://jetp.ras.ru/cgi-bin/e/index/r/49/1/p345?a=list}}].


\bibitem{bardeen}
J. Bardeen, presented at GR5, Tiflis, U.S.S.R., and published in the conference proceedings in the U.S.S.R. (1968).

\bibitem{39}
Eloy Ayón-Beato, Alberto García, {\it{Regular Black Hole in General Relativity Coupled to Nonlinear Electrodynamics}}, Phys. Rev. Lett. {\bf 80}: 5056-5059, (1998), 
[\href{http://arxiv.org/abs/gr-qc/9911046}{{\tt   	gr-qc/9911046}}];

Kirill A. Bronnikov, {\it{Regular Magnetic Black Holes and Monopoles from Nonlinear Electrodynamics}}, Phys.Rev.D {63}: 044005, (2001), 
[\href{http://arxiv.org/abs/gr-qc/0006014}{{\tt  gr-qc/0006014}}];

Irina Dymnikova, {\it{Regular electrically charged structures in Nonlinear Electrodynamics coupled to General Relativity}}, Class.Quant.Grav {\bf .21}: 4417-4429, (2004), 
[\href{http://arxiv.org/abs/gr-qc/0407072}{{\tt  	gr-qc/0407072}}];

M. Novello, V. A. De Lorenci, J. M. Salim, R. Klippert, {\it{Geometrical aspects of light propagation in nonlinear electrodynamics}}, Phys.Rev. D {\bf 61} (2000) 045001, [\href{http://arxiv.org/abs/gr-qc/9911085}{{\tt  gr-qc/9911085}}];

\bibitem{HEllis}
S. W. Hawking and G.F.R. Ellis, {\it{The Large Scale Struc-
ture of Space-Time}}, (Cambridge University Press, Cambridge, 1973).

\bibitem{Wormholes}
M. S. Morris and K. S. Thorne, {\it{Wormholes in space-time and their use for interstellar travel: A tool for teaching general
relativity}}, Am. J. Phys. {\bf 56} (1988), 395-412 [\href{https://aapt.scitation.org/doi/10.1119/1.15620}{{\tt DOI: 10.1119/1.15620}}];

M. S. Morris, K. S. Thorne and U. Yurtsever, {\it{Wormholes, Time Machines, and the Weak Energy Condition}}, Phys.
Rev. Lett. {\bf 61} (1988), 1446-1449 [\href{https://journals.aps.org/prl/abstract/10.1103/PhysRevLett.61.1446}{{\tt https://doi.org/10.1103/PhysRevLett.61.1446}}];

\bibitem{Visser}
M. Visser,{\it{ Lorentzian wormholes: From Einstein to Hawking}}, AIP press [now Springer], New York (1995).
\bibitem{LOBO}
] F. S. N. Lobo, Wormholes, Warp Drives and Energy Conditions, Fundam. Theor. Phys. {\bf 189}, pp. (2017), (formerly
Lecture Notes in Physics), Springer Nature Switzerland AG.


\bibitem{Visser2}
Alex Simpson and Matt Visser, {\it{Black-bounce to traversable wormhole}}, JCAP 02(2019)042,[\href{https://doi.org/10.1088/1475-7516/2019/02/042}{{\tt DOI:10.1088 / 1475-7516 / 2019/02/042}}].

\bibitem{manuelmarcos}
Francisco S. N. Lobo, Manuel E. Rodrigues, Marcos V. de S. Silva, Alex Simpson and Matt Visser, {\it{Novel black-bounce spacetimes: wormholes, regularity, energy conditions, and causal structure}}, Phys. Rev. D {\bf 103}, 084052 (2021), [\href{https://doi.org/10.1103/PhysRevD.103.084052}{{\tt DOI: 10.1103/PhysRevD.103.084052}}].

\bibitem{Visser3}
A. Simpson, P. Martin Moruno and M. Visser, {\it Vaidya spacetimas, black-bounce, and traversable wormholes}, Class.
Quant. Grav. 36 (2019) no.14, 145007 [\href{https://arxiv.org/abs/1902.04232}{{\tt 	arXiv:1902.04232 [gr-qc]}}].

\bibitem{Visser4}
F. S. N. Lobo, A. Simpson and M. Visser, {\it Dynamic thin-shell black-bounce traversable wormholes}, Phys. Rev. D 101
(2020) no.12, 124035 [\href{https://arxiv.org/abs/2003.09419}{{\tt arXiv:2003.09419 [gr-qc]}}].




%
%



%

%

%

%

%
\bibitem{Ia}
A. Riess et al, {\it{Observational Evidence from Supernovae for an Accelerating Universe and a Cosmological Constant}}, Astron. J. {\bf 116}, 1009 (1998) [\href{https://arxiv.org/abs/astro-ph/9805201}{{\tt 	arXiv:astro-ph/9805201}}];
 S. Perlmutter et al, {\it{Measurements of Omega and Lambda from 42 High-Redshift Supernovae
}}, Astrophys. J. {\bf 517}, 565 (1999) [\href{https://arxiv.org/abs/astro-ph/9812133}{{\tt 	arXiv:astro-ph/9812133}}]; J. Tonry et al, {\it Cosmological Results from High-z Supernovae}, Astrophys.J. {\bf 594}, 1 (2003) [\href{https://arxiv.org/abs/astro-ph/0305008}{{\tt 	arXiv:astro-ph/0305008}}]. 
\bibitem{Starobinski}
A. A. Starobinsky, {\it{A new type of isotropic cosmological
models without singularity}}, Phys. Lett. B {\bf 91}, 99 (1980) [\href{https://doi.org/10.1016/0370-2693(80)90670-X}{{\tt 10.1016/0370-2693(80)90670-X}}];
%
\bibitem{fR}
S. Nojiri, S.D. Odintsov, {\it{Introduction to Modified Gravity and Gravitational Alternative for Dark Energy}}, ECONF C0602061, {\bf 06}, (2006); Int. J. Geom. Meth. Mod.Phys. 4,115-146,2007 [\href{http://arxiv.org/abs/hep-th/0601213}{{\tt hep-th/0601213}}];
Thomas P. Sotiriou, Valerio Faraoni, {\it{f(R) Theories Of Gravity}}, Rev. Mod. Phys. {\bf 82} 451-497 (2010), [\href{http://arxiv.org/abs/0805.1726}{{\tt  	arXiv:0805.1726}}];
Antonio De Felice, Shinji Tsujikawa, {\it{f(R) theories}}, Living Rev. Rel. {\bf 13} 3, (2010), [\href{http://arxiv.org/abs/1002.4928}{{\tt  arXiv:1002.4928}}];
Timothy Clifton, Pedro G. Ferreira, Antonio Padilla, Constantinos Skordis, {\it{Modified Gravity and Cosmology}}, Physics Reports {\bf 513}, 1 (2012), 1-189, [\href{http://arxiv.org/abs/1106.2476}{{\tt  	arXiv:1106.2476}}].
%

%
 
%
\bibitem{fRT}
F. G. Alvarenga, M. J. S. Houndjo, A. V. Monwanou, Jean B. Chabi Orou, {\it{Testing some f(R,T) gravity models from energy conditions}}, Journal of Modern Physics, {\bf 4}, 130-139 (2013), [\href{http://arxiv.org/abs/1205.4678}{{\tt  arXiv:1205.4678}}];
M. J. S. Houndjo, C. E. M. Batista, J. P. Campos, O. F. Piattella, {\it{Finite-time singularities in f(R, T) gravity and the effect of conformal anomaly}}, Can. J. Phys. {\bf 91}(7), 548-553 (2013), [\href{http://arxiv.org/abs/1203.6084}{{\tt  arXiv:1203.6084 }}];
Mubasher Jamil, D. Momeni, Muhammad Raza, Ratbay Myrzakulov, {\it{Reconstruction of some cosmological models in f(R,T) gravity}}, Eur. Phys. J. C {\bf 72} 1999 (2012), [\href{http://arxiv.org/abs/1107.5807}{{\tt  	arXiv:1107.5807}}];
M. J. S. Houndjo, Oliver F. Piattella, {\it{Reconstructing f(R,T) gravity from holographic dark energy}}, IJMPD {\bf2}, 1250024 (2012), [\href{http://arxiv.org/abs/1111.4275}{{\tt  	arXiv:1111.4275}}];
M. J. S. Houndjo, {\it{Reconstruction of f(R, T) gravity describing matter dominated and accelerated phases}},  	IJMPD {\bf 21}, 1250003 (2012), [\href{http://arxiv.org/abs/1107.3887}{{\tt  	arXiv:1107.3887}}];
Tiberiu Harko, Francisco S.N. Lobo, Shin'ichi Nojiri, Sergei D. Odintsov, {\it{f(R,T) gravity}}, Phys.Rev.D {\bf84} 024020, (2011), [\href{http://arxiv.org/abs/1104.2669}{{\tt  	arXiv:1104.2669}}].

%
%
\bibitem{fG}
Kazuharu Bamba, Chao-Qiang Geng, Shin'ichi Nojiri, Sergei D. Odintsov, {\it{Equivalence of modified gravity equation to the Clausius relation}}, Europhys. Lett. {\bf 89} (2010) 50003, [\href{http://arxiv.org/abs/0909.4397}{{\tt  arXiv:0909.4397}}]; 
M. J. S. Houndjo, M. E. Rodrigues, D. Momeni, R. Myrzakulov, {\it{Exploring Cylindrical Solutions in Modified f(G) Gravity}}, Canadian Journal of Physics, 2014,  {\bf 92}(12) 1528-1540, [\href{http://arxiv.org/abs/1301.4642}{{\tt  arXiv:1301.4642}}];
M. E. Rodrigues, M. J. S. Houndjo, D. Momeni, R. Myrzakulov, {\it{ A Type of Levi-Civita's Solution in Modified Gauss-Bonnet Gravity}}, Canadian Journal of Physics, 2014, {\bf92}(2): 173-176, [\href{http://arxiv.org/abs/1212.4488}{{\tt  arXiv:1212.4488}}];
Kazuharu Bamba, Sergei D. Odintsov, Lorenzo Sebastiani, Sergio Zerbini, {\it{Finite-time future singularities in modified Gauss-Bonnet and F(R,G) gravity and singularity avoidance}}, Eur.Phys.J.C{\bf 67} 295-310, (2010), [\href{http://arxiv.org/abs/0911.4390}{{\tt  	arXiv:0911.4390}}];
Shin'ichi Nojiri, Sergei D. Odintsov, Alexey Toporensky, Petr Tretyakov, {\it{Reconstruction and deceleration-acceleration transitions in modified gravity}}, Gen.Rel.Grav.{\bf 42} 1997-2008 , (2010), [\href{http://arxiv.org/abs/0912.2488}{{\tt arXiv:0912.2488}}].
%
\bibitem{35marcos}
M. E. Rodrigues, E. L. B. Junior, G. T. Marques and V. T. Zanchin, {\it{Regular black holes in f (R) gravity coupled to
nonlinear electrodynamics}}, Phys. Rev. D {\bf 94} (2016) no.2, 024062  [\href{https://arxiv.org/abs/1511.00569}{{\tt arXiv:1511.00569 [gr-qc]}}];

L. Hollenstein and F. S. N. Lobo, {\it{Exact solutions of f (R) gravity coupled to nonlinear electrodynamics}}, Phys. Rev. D {\bf 78} (2008), 124007, [\href{https://arxiv.org/abs/0807.2325}{{\tt arXiv:0807.2325 [gr-qc]}}];

M. Guerrero and D. Rubiera-Garcia, {\it{Nonsingular black holes in nonlinear gravity coupled to Euler-Heisenberg electrodynamics}}, Phys. Rev. D {\bf 102} (2020) no.2, 024005 [\href{https://arxiv.org/abs/2005.08828}{{\tt arXiv:2005.08828 [gr-qc]}}].

\bibitem{29}
O. B. Zaslavskii, {\it {Regular black holes and energy conditions}}, Phys. Lett. B {\bf 688}, 278-280 (2010) [\href{https://arxiv.org/abs/1004.2362}{{\tt arXiv:1004.2362 [gr-qc]}}].

\bibitem{47}
M. E. Rodrigues, J. C. Fabris, E. L. B. Junior and G. T. Marques, {\it {Generalisation for regular black holes on general
relativity to f (R) gravity}}, Eur. Phys. J. C {\bf 76} (2016) no.5, 250, [\ref{https://arxiv.org/abs/1601.00471}{{\tt arXiv:1601.00471 [gr-qc]}}].

\bibitem{48}
M. E. Rodrigues, E. L. B. Junior and M. V. de S. Silva, {\it {Using dominant and weak energy conditions for building new
classes of regular black holes}}, JCAP {\bf 02} (2018), 059, [\href{https://arxiv.org/abs/1705.05744}{{\tt arXiv:1705.05744 [physics.gen-ph]}}].

\bibitem{49}
M. E. Rodrigues and M. V. de S. Silva, {\it {Bardeen Regular Black Hole With an Electric Source}}, JCAP {\bf 06} (2018), 025
[\href{https://arxiv.org/abs/1802.05095}{{\tt arXiv:1802.05095 [gr-qc]}}]; 

C. Bambi, L. Modesto, {\it {Rotating regular black holes}}, Phys. Lett. B {\bf 721} (2013), 329-334, [\href{https://arxiv.org/abs/1302.6075}{{\tt arXiv:1302.6075 [gr-qc]}}];

J. C. S. Neves, A. Saa, {\it {Regular rotating black holes and the weak energy condition}}, Phys. Lett. B {\bf 734} (2014), 44-48,
[\href{https://arxiv.org/abs/1402.2694}{{\tt arXiv:1402.2694 [gr-qc]}}];

B. Toshmatov, B. Ahmedov, A. Abdujabbarov, Z. Stuchlik, {\it{Rotating Regular Black Hole Solution}}, Phys. Rev. D {\bf 89}
(2014) no. 10, 104017, [\href{https://arxiv.org/abs/1404.6443}{{\tt arXiv:1404.6443 [gr-qc]}}];

M. Azreg-Ainou, {\it {Generating rotating regular black hole solutions without complexification}}, Phys. Rev. D {\bf 90} (2014)
no. 6, 064041, [\href{https://arxiv.org/abs/1405.2569}{{\tt arXiv:1405.2569 [gr-qc]}}];

I. Dymnikova, E. Galaktionov, {\it{Regular rotating electrically charged black holes and solitons in non-linear electrodynamics
minimally coupled to gravity}}, Class. Quant. Grav. {\bf 32} (2015) no. 16, 165015, [\href{https://arxiv.org/abs/1510.01353}{{\tt arXiv:1510.01353 [gr-qc]}}];

R. Torres, F. Fayos, {\it{On regular rotating black holes}}, Gen. Rel. Grav. {\bf 49} (2017) no. 1, 2, Quant. Grav. 32 (2015) no.
16, 165015, [\href{https://arxiv.org/abs/1611.03654}{{\tt arXiv:1611.03654 [gr-qc]}}].

\bibitem{TT}
R. Aldrovandi and J. G. Pereira, {\it{An Introduction to Teleparallel Gravity}}, Instituto de Fisica Teorica, UNSEP, Sao Paulo, [\href{www.ift.unesp.br/users/jpereira/tele.pdf}{{\tt www.ift.unesp.br/users/jpereira/tele.pdf}}];
R. Aldrovandi; J. G. Pereira; K. H. Vu, {\it{Selected topics in teleparallel gravity }}, Braz. J. Phys. vol.{\bf 34} no.4a São Paulo Dec. 2004, [\href{http://arxiv.org/abs/gr-qc/0312008}{{\tt gr-qc/0312008}}]; J.W. Maluf, {\it{The teleparallel equivalent of general relativity}}, Annalen Phys. {\bf 525} (2013) 339-357, [\href{http://arxiv.org/abs/arXiv:1303.3897}{{\tt arXiv:1303.3897}}];  F. W. Hehl, J. D. McCrea, E. W. Mielke, and Y. Ne'eman, {\it{Metric-affine gauge theory of gravity: field equations, Noether identities, world spinors, and breaking of dilation invariance}}, Phys. Rep. {\bf 258}, 1-171 (1995), [\href{http://www.sciencedirect.com/science/article/pii/037015739400111F}{{\tt DOI:10.1016/0370-1573(94)00111-F}}].
%

%
%


%

\bibitem{fT}
Tiberiu Harko, Francisco S. N. Lobo, G. Otalora, Emmanuel N. Saridakis, {\it{Nonminimal torsion-matter coupling extension of f(T) gravity }}, Phys.Rev. D {\bf89} (2014) 124036,
[\href{http://arxiv.org/abs/arXiv:1404.6212}{{\tt arXiv:1404.6212}}]; S. Basilakos, S. Capozziello, M. De Laurentis, A. Paliathanasis, M. Tsamparlis, {\it{Noether symmetries and analytical solutions in f(T)-cosmology: A complete study }}, Phys.Rev. D {\bf88} (2013) 103526, 
[\href{http://arxiv.org/abs/arXiv:1311.2173}{{\tt arXiv:1311.2173}}]; Kazuharu Bamba, Sergei D. Odintsov, Diego Sáez-Gómez, {\it{Conformal symmetry and accelerating cosmology in teleparallel gravity }}, Phys.Rev. D {\bf88} (2013) 084042,
[\href{http://arxiv.org/abs/arXiv:1308.5789}{{\tt arXiv:1308.5789}}]; H. Mohseni Sadjadi, {\it{Generalized Noether symmetry in f(T) gravity}}, Phys.Lett. B {\bf718} (2012) 270-275, 
[\href{http://arxiv.org/abs/arXiv:1210.0937}{{\tt arXiv:1210.0937}}]; 
M.E. Rodrigues, M.J.S. Houndjo, D. Saez-Gomez, F. Rahaman, {\it{Anisotropic Universe Models in f(T) Gravity}}, Phys.Rev. D {\bf86} (2012) 104059,
[\href{http://arxiv.org/abs/arXiv:1209.4859}{{\tt http://arxiv.org/abs/arXiv:1209.4859}}]; Vincenzo F. Cardone, Ninfa Radicella, Stefano Camera, {\it{Accelerating f(T) gravity models constrained by recent cosmological data }}, Phys.Rev. D {\bf85} (2012) 124007,
[\href{http://arxiv.org/abs/arXiv:1204.5294}{{\tt arXiv:1204.5294}}]; Kazuharu Bamba, Ratbay Myrzakulov, Shin'ichi Nojiri, Sergei D. Odintsov, {\it{Reconstruction of f(T) gravity: Rip cosmology, finite-time future singularities and thermodynamics}},  Phys.Rev. D {\bf85} (2012) 104036, 
[\href{http://arxiv.org/abs/arXiv:1202.4057}{{\tt arXiv:1202.4057}}]; Chen Xu, Emmanuel N. Saridakis, Genly Leon, {\it{Phase-Space analysis of Teleparallel Dark Energy}}, JCAP {\bf1207} (2012) 005, 
[\href{http://arxiv.org/abs/arXiv:1202.3781}{{\tt arXiv:1202.3781}}]; K. Karami, A. Abdolmaleki, {\it{Generalized second law of thermodynamics in f(T)-gravity }}, JCAP {\bf1204} (2012) 007, 
[\href{http://arxiv.org/abs/arXiv:1201.2511}{{\tt arXiv:1201.2511}}]; 
Hao Wei, Xiao-Jiao Guo, Long-Fei Wang, {\it{Noether Symmetry in f(T) Theory}},  Phys.Lett. B707 (2012) 298-304, 
[\href{http://arxiv.org/abs/arXiv:1112.2270}{{\tt arXiv:1112.2270}}]; Christian G. Boehmer, Tiberiu Harko, Francisco S.N. Lobo, {\it{ Wormhole geometries in modified teleparralel gravity and the energy conditions }}, Phys.Rev. D {\bf85} (2012) 044033,
[\href{http://arxiv.org/abs/arXiv:1110.5756}{{\tt arXiv:1110.5756}}]; Yi-Peng Wu, Chao-Qiang Geng, {\it{Primordial Fluctuations within Teleparallelism }}, Phys.Rev. D {\bf86} (2012) 104058, 
[\href{http://arxiv.org/abs/arXiv:1110.3099}{{\tt arXiv:1110.3099}}]; S. Capozziello, V.F. Cardone, H. Farajollahi, A. Ravanpak, {\it{ Cosmography in f(T)-gravity }}, Phys.Rev. D {\bf84} (2011) 043527, 
[\href{http://arxiv.org/abs/arXiv:1108.2789}{{\tt arXiv:1108.2789}}]; Christian G. Boehmer, Atifah Mussa, Nicola Tamanini, {\it{Existence of relativistic stars in f(T) gravity}}, Class.Quant.Grav. {\bf28} (2011) 245020, 
[\href{http://arxiv.org/abs/arXiv:1107.4455}{{\tt arXiv:1107.4455}}]; Rong-Xin Miao, Miao Li, Yan-Gang Miao, {\it{Violation of the first law of black hole thermodynamics in f(T) gravity}},  JCAP {\bf1111} (2011) 033, 
[\href{http://arxiv.org/abs/arXiv:1107.0515}{{\tt arXiv:1107.0515}}]; Xin-he Meng, Ying-bin Wang, {\it{Birkhoff's theorem in the f(T) gravity }}, Eur.Phys.J. C {\bf71} (2011) 1755, 
[\href{http://arxiv.org/abs/arXiv:1107.0629}{{\tt arXiv:1107.0629}}]; Hao Wei, Xiao-Peng Ma, Hao-Yu Qi, {\it{f(T) Theories and Varying Fine Structure Constant}}, Phys.Lett. B {\bf703} (2011) 74-80, 
[\href{http://arxiv.org/abs/arXiv:1106.0102}{{\tt arXiv:1106.0102}}]; Miao Li, Rong-Xin Miao, Yan-Gang Miao, {\it{Degrees of freedom of f(T) gravity }}, JHEP {\bf1107} (2011) 108, 
[\href{http://arxiv.org/abs/arXiv:1105.5934}{{\tt arXiv:1105.5934}}]; Yi-Fu Cai, Shih-Hung Chen, James B. Dent, Sourish Dutta, Emmanuel N. Saridakis, {\it{Matter Bounce Cosmology with the f(T) Gravity }}, Class.Quant.Grav. {\bf28} (2011) 215011
[\href{http://arxiv.org/abs/arXiv:1104.4349}{{\tt arXiv:1104.4349}}]; Rafael Ferraro, Franco Fiorini, {\it{Non trivial frames for f(T) theories of gravity and beyond }}, Phys.Lett. B {\bf702} (2011) 75-80, 
[\href{http://arxiv.org/abs/arXiv:1103.0824}{{\tt arXiv:1103.0824}}]; Yi Zhang, Hui Li, Yungui Gong, Zong-Hong Zhu, {\it{Notes on f(T) Theories }}, JCAP {\bf1107} (2011) 015, 
[\href{http://arxiv.org/abs/arXiv:1103.0719}{{\tt arXiv:1103.0719}}]; Tower Wang, {\it{Static Solutions with Spherical Symmetry in f(T) Theories}}, Phys.Rev. D {\bf84} (2011) 024042, 
[\href{http://arxiv.org/abs/arXiv:1102.4410}{{\tt arXiv:1102.4410}}]; Thomas P. Sotiriou, Baojiu Li, John D. Barrow, {\it{Generalizations of teleparallel gravity and local Lorentz symmetry }}, Phys.Rev. D {\bf83} (2011) 104030, 
[\href{http://arxiv.org/abs/arXiv:1012.4039}{{\tt arXiv:1012.4039}}]; Kazuharu Bamba, Chao-Qiang Geng, Chung-Chi Lee, Ling-Wei Luo, {\it{ Equation of state for dark energy in f(T) gravity }}, JCAP {\bf1101} (2011) 021, 
[\href{http://arxiv.org/abs/arXiv:1011.0508}{{\tt arXiv:1011.0508}}]; Rui Zheng, Qing-Guo Huang, {\it{Growth factor in f(T) gravity }}, JCAP {\bf1103} (2011) 002, 
[\href{http://arxiv.org/abs/arXiv:1010.3512}{{\tt arXiv:1010.3512}}]; James B. Dent, Sourish Dutta, Emmanuel N. Saridakis, {\it{f(T) gravity mimicking dynamical dark energy. Background and perturbation analysis }}, JCAP {\bf1101} (2011) 009, 
[\href{http://arxiv.org/abs/arXiv:1010.2215}{{\tt arXiv:1010.2215}}];  Rong-Jia Yang, {\it{Conformal transformation in f(T) theories}}, Europhys.Lett. {\bf93} (2011) 60001, 
[\href{http://arxiv.org/abs/arXiv:1010.1376}{{\tt arXiv:1010.1376}}]; Baojiu Li, Thomas P. Sotiriou, John D. Barrow, {\it{f(T) gravity and local Lorentz invariance }}, Phys.Rev. D {\bf83} (2011) 064035, 
[\href{http://arxiv.org/abs/arXiv:1010.1041}{{\tt arXiv:1010.1041}}];  K. Karami, A. Abdolmaleki, {\it{f(T) modified teleparallel gravity models as an alternative for holographic and new agegraphic dark energy models }}, Res.Astron.Astrophys. {\bf13} (2013) 757-771, 
[\href{http://arxiv.org/abs/arXiv:1009.2459}{{\tt arXiv:1009.2459}}]; Puxun Wu, Hong Wei Yu, {\it{f(T) models with phantom divide line crossing}}, Eur.Phys.J. C {\bf71} (2011) 1552, 
[\href{http://arxiv.org/abs/arXiv:1008.3669}{{\tt arXiv:1008.3669}}]; Gabriel R. Bengochea, {\it{Observational information for f(T) theories and Dark Torsion }}, Phys.Lett. B {\bf695} (2011) 405-411,
[\href{http://arxiv.org/abs/arXiv:1008.3188}{{\tt arXiv:1008.3188}}]; Shih-Hung Chen, James B. Dent, Sourish Dutta, Emmanuel N. Saridakis, {\it{Cosmological perturbations in f(T) gravity }}, Phys.Rev. D {\bf83} (2011) 023508, 
[\href{http://arxiv.org/abs/arXiv:1008.1250}{{\tt arXiv:1008.1250}}]; Rong-Jia Yang, {\it{New types of f(T) gravity }}, Eur.Phys.J. C {\bf71} (2011) 179, 
[\href{http://arxiv.org/abs/arXiv:1007.3571}{{\tt arXiv:1007.3571}}]; Puxun Wu, Hong Wei Yu, {\it{The dynamical behavior of f(T) theory }}, Phys.Lett. B {\bf692} (2010) 176-179, 
[\href{http://arxiv.org/abs/arXiv:1007.2348}{{\tt arXiv:1007.2348}}]; Ratbay Myrzakulov, {\it{Accelerating universe from F(T) gravity }}, Eur.Phys.J. C {\bf71} (2011) 1752, 
[\href{http://arxiv.org/abs/arXiv:1006.1120}{{\tt arXiv:1006.1120}}]; G.G.L. Nashed and W. El Hanafy, {\it{A Built-in Inflation in the f(T)-Cosmology}}, Eur.Phys.J. C {\bf 74} (2014) 10, 3099, 
[\href{http://arxiv.org/abs/arXiv:1403.0913}{{\tt arXiv:1403.0913}}].
%

\bibitem{ferraro1}
Rafael Ferraro and Franco Fiorini, {\it{Modified teleparallel gravity: Inflation without inflaton }},  Phys.Rev. D {\bf 75} (2007) 084031, 
      [\href{http://arxiv.org/abs/gr-qc/0610067}
{{\tt gr-qc/0610067}}].


\bibitem{ednaldo}
Ednaldo LB Junior , Manuel E. Rodrigues , Mahouton JS Houndjo, {\it{Born-Infeld and Charged Black Holes with non-linear source in f(T)}}, JCAP {\bf 06} (2015)037 [\href{https://arxiv.org/abs/1503.07427}{{\tt 	arXiv:1503.07427 [gr-qc]}}].

\bibitem{rodrigues4}
Ednaldo L. B. Junior, Manuel E. Rodrigues and Mahouton J. S. Houndjo, {\it{Regular black holes in f(T) Gravity through a nonlinear electrodynamics source}}, JCAP {\bf 10} (2015) 060 [\href{https://arxiv.org/abs/1503.07857}{{\tt arXiv:1503.07857 [gr-qc]}}].

\bibitem{reboucas}
Di Liu, M. J. Reboucas, {\it{Energy conditions bounds on f(T) gravity}}, 	Phys.Rev. D {\bf 86} (2012) 083515 [\href{https://arxiv.org/abs/1207.1503}{{\tt arXiv:1207.1503 [astro-ph.CO]}}].

\bibitem{fernando1}
Sharmanthie Fernando and Don Krug, {\it{Charged black hole solutions in Einstein-Born-Infeld gravity with a cosmological constant }},  Gen.Rel.Grav. {\bf 35} (2003) 129-137, 
      [\href{http://arxiv.org/abs/hep-th/0306120}
{{\tt hep-th/0306120}}].
%

\bibitem{rodrigues1}
M.E. Rodrigues, M.J.S. Houndjo, J. Tossa, D. Momeni and R. Myrzakulov, {\it{Charged Black Holes in Generalized Teleparallel Gravity }},  JCAP {\bf 1311} (2013) 024, 
      [\href{http://arxiv.org/abs/arXiv:1306.2280}
{{\tt arXiv:1306.2280}}].
%
\bibitem{rodrigues2}
Manuel E. Rodrigues, M. Hamani Daouda and  M. J. S. Houndjo, {\it{Inhomogeneous Universe in f(T) Theory }},  Grav.Cosmol. {\bf 20} (2014) 2, 80-89, 
      [\href{http://arxiv.org/abs/arXiv:1205.0565}
{{\tt arXiv:1205.0565}}].
%
\bibitem{plebanski}
J. Plebanski and A. Krasinski, {\it{An Introduction to General Relativity and Cosmology}}, Cambridge University Press, New York (2006). 
%

\bibitem{capozziello1}
Salvatore Capozziello, P.A. Gonzalez, Emmanuel N. Saridakis, Yerko Vasquez, {\it{Exact charged black-hole solutions in D-dimensional f(T) gravity: torsion vs curvature analysis}}, JHEP {\bf1302} (2013) 039,
[\href{http://arxiv.org/abs/arXiv:1210.1098}{{\tt arXiv:1210.1098}}].
%
\bibitem{aldrovandi}
Ruben Aldrovandi and Jose Geraldo Pereira, {\it{Teleparallel Gravity, An Introduction}}, Springer, New York (2013). 
%
\bibitem{tamanini}
Nicola Tamanini, Christian G. Boehmer, {\it{Good and bad tetrads in f(T) gravity}}, Phys.Rev. D {\bf86} (2012) 044009, 
[\href{http://arxiv.org/abs/arXiv:1204.4593}{{\tt arXiv:1204.4593}}].
%
\bibitem{rodrigues3}
M. H. Daouda, M. E. Rodrigues and M.J.S. Houndjo, {\it{Anisotropic fluid for a set of non-diagonal tetrads in f(T) gravity}}, Phys.Lett. B {\bf 715} (2012) 241-245, 
[\href{http://arxiv.org/abs/arXiv:1202.1147}{{\tt arXiv:1202.1147}}]. 
%
\bibitem{wainwright}
J. Wainwright and P. E. A. Yaremovicz,  {\it{Killing vector fields and the Einstein-Maxwell field equations with perfect fluid source}}, Gen. Rel. Grav. {\bf 7}, 345 (1976) [\href{https://link.springer.com/article/10.1007/BF00771105}{{\tt DOI:10.1007/BF00771105}}]; J. Wainwright, P.E.A Yaremovicz, {\it{Symmetries
of the Einstein-Maxwell field equations: the null field case}}, Gen. Rel. Grav. {\bf 7}, 595 (1976) [\href{https://link.springer.com/article/10.1007/BF00763408}{{\tt DOI :10.1007/BF00763408}}].
%
\bibitem{hollenstein}
Lukas Hollenstein and Francisco S. N. Lobo,  {\it{Exact solutions of f(R) gravity coupled to nonlinear electrodynamics}}, XXX, 
[\href{http://arxiv.org/abs/0807.2325v2}{{\tt arXiv:0807.2325}}]. 
%

%


\end{thebibliography}
\end{document}